 \documentclass[final,3p,times]{elsarticle}
\usepackage{amssymb}
\usepackage{amsmath}
\usepackage{amsfonts}
\usepackage{appendix}
\usepackage{bm}
\usepackage[caption=false,subrefformat=parens]{subfig}
\usepackage{graphicx}
\usepackage{epsfig}
\usepackage{epstopdf}
\usepackage{balance}
\usepackage[dvipsnames]{xcolor}
\usepackage{calc}
\usepackage[colorlinks,
linkcolor=blue,
anchorcolor=blue,
citecolor=blue,
urlcolor=blue]{hyperref}
\usepackage{lipsum}
\usepackage{braket}
\usepackage{subfig}
\usepackage{verbatim}
\usepackage{amsthm}

\journal{xxx}

\begin{document}

\begin{frontmatter}



\title{Discrete unified gas kinetic scheme for the solution of electron Boltzmann transport equation with Callaway approximation}% Force line breaks with \\



\author[a]{Meng Lian}

\affiliation[a]{organization={School of Physics, Institute for Quantum Science and Engineering and Wuhan National High Magnetic Field Center, Huazhong University of Science and Technology},
            city={Wuhan 430074},
            country={China}}

%\affiliation[a]{School of Physics, Institute for Quantum Science and Engineering and Wuhan National High Magnetic Field Center,
% Huazhong University of Science and Technology, Wuhan 430074,  China}

\author[b]{Chuang Zhang}

\affiliation[b]{organization={Department of Physics, Hangzhou Dianzi University},
            city={Hangzhou 310018},
            country={China}}

%\affiliation[b]{Department of Physics, Hangzhou Dianzi University, Hangzhou 310018, China}

 \author[d]{Zhaoli Guo\corref{cor1}}%
 \ead{zlguo@hust.edu.cn}
% \affiliation[c]{State Key Laboratory of Coal Combustion, Huazhong University of Science and Technology, Wuhan, 430074, China}
\affiliation[d]{organization={Institute of Interdisciplinary Research for Mathematics and Applied Science, Huazhong University of Science and Technology},
            city={Wuhan 430074},
            country={China}}
 %\affiliation[d]{Institute of Interdisciplinary Research for Mathematics and Applied Science, Huazhong University of Science and Technology, Wuhan, 430074, China}

\author[a]{Jing-Tao Lü\corref{cor1}}%
 \ead{jtlu@hust.edu.cn}
 
 \cortext[cor1]{Corresponding author}
% \affiliation{School of Physics, Institute for Quantum Science and Engineering and Wuhan National High Magnetic Field Center,
%Huazhong University of Science and Technology, Wuhan 430074,  China}%Lines break automatically or can be forced with \\

\begin{abstract}
Electrons are the carriers of heat and electricity in materials, and exhibit abundant transport phenomena such as ballistic, diffusive, and hydrodynamic behaviors in systems with different sizes. The electron Boltzmann transport equation (eBTE) is a reliable model for describing electron transport, but it is a challenging problem to efficiently obtain the numerical solutions of eBTE within one unified scheme involving ballistic, hydrodynamics and/or diffusive regimes. In this work, a discrete unified gas kinetic scheme (DUGKS) in finite-volume framework is developed based on the eBTE with the Callaway relaxation model for electron transport. By reconstructing the distribution function at the cell interface, the processes of electron drift and scattering are coupled together within a single time step. Numerical tests demonstrate that the DUGKS can be adaptively applied to multiscale electron transport, across different regimes.

\end{abstract}




\end{frontmatter}


\section{Introduction}

With the development of nanoscience, microscale processing technologies, and advancements in nanoscale temperature measurement techniques, especially the continuous miniaturization and integration of chips, the carrier transport behavior of nano systems has become an important scientific issue in the statistical physics of nonequilibrium states \cite{10.1063/1.4832615,Cahill2003793,10.1063/1.5062841,RevModPhys.85.1295,aam9991}. For transport in metals and semiconductors, electrons play an important role and are characterized by rich transport phenomena, such as thermoelectric effect \cite{10.1063/1.3060810,bulusu_review_2008,Ren_2023}, hydrodynamics \cite{levitov_electron_2016,aharon-steinberg_direct_2022,kumar_imaging_2022}, and size effects \cite{book2}. As one of the widely used methods for studying electron transport, the electron Boltzmann equation (eBTE) can be used to solve problems over a large range of scales from the nanoscale to the macroscale \cite{2005Nanoscale}. The Callaway approximation \cite{PhysRev.113.1046} divides the scattering processes into two types. One conserves the total crystal momenta, while the other does not.  It is able to successfully characterize transport behavior in the hydrodynamic regime \cite{PhysRevB.99.165409,PhysRevLett.126.076803}, which has received renewed interest in recent years, due to the experimental progress in graphene and other two-dimensional materials \cite{doi:10.1126/science.aad0201,moll_evidence_2016,doi:10.1126/science.aad0343}.  Development of  multiscale and high accuracy method for solving eBTE under the Callaway approximation is highly desirable to further explore the complex characteristics of electron transport in realistic materials.

The numerical methods of eBTE are mainly divided into two types. One is to use macroscopic models to approximate the solution, such as the drift-diffusion model and the hydrodynamic equations \cite{https://doi.org/10.1002/pssb.19690350206,1476105,1477063}. These methods are simple in form, with high computational efficiency, and can accurately describe the electron transport behavior in large-scale devices. However, as the size of the device decreases, the electron mean free path is comparable to the characteristic length of the device. In such cases, the electron transport behavior exhibits quasi-ballistic or ballistic characteristics, rendering the aforementioned methods inaccurate.
The other type is the direct numerical solution of the eBTE, such as the Monte Carlo (MC) method, the discrete ordinates method (DOM) and the spherical harmonic expansion (SHE) method. The MC method uses random numbers to select the scattering mechanisms and determine the drift time under external field. It is very flexible and can be combined with the actual energy band structure to analyze the complex scattering process of the carriers. Currently, MC schemes have been developed for semiconductors \cite{RevModPhys.55.645} and metals \cite{PhysRevB.99.205433}, but are not suitable for weak external fields due to unavoidable statistical noise, i.e., it converges slowly at small Knudsen numbers due to decoupling of particle drift and scattering. The DOM method discretizes both real space and wavevector space, and solves the eBTE in the whole real space for each discretized wavevector \cite{doi:10.1080/10407799308955899,huang_conservative_2015}. This method is suitable for systems with high Knudsen number, but exhibits significant numerical dissipation at small Knudsen number. The SHE method is widely used for the simulation of semiconductor devices \cite{goldsman_physics-based_1991,gnudi1991one,541439,hennacy_deterministic_1995}. By representing the electron distribution with spherical harmonic functions of a certain order, a series of equations are obtained, which can accurately describe the transport behavior of carriers. It is more advantageous than MC in weak external fields, and does not suffer from statistical noise. However, the numerical results depend on the order of the expansion, which needs to be taken care of in the high-field region. Meanwhile, the high computational complexity of SHE is challenging to simulate high-dimensional materials, which requires the use of matrix compression techniques \cite{rupp_matrix_2010}. In addition, the lattice Boltzmann methods (LBM) \cite{guo_lattice_2013} have been applied to multiscale problems in nanosystems, but it may generate nonphysical solutions at high Knudsen numbers \cite{doi:10.1080/10407790.2014.915683}.

%For solving the BTE several methods have been developed: the discrete ordinates method (DOM) method is adopted widely \cite{doi:10.1080/10407799308955899,huang_conservative_2015}, but due to the decoupled treatment of particle advection and scattering, the cell size and time step are required to be smaller than the mean free path and relaxation time , which is suitable for the high Knudsen number case; in the Monte Carlo (MC) method, schemes for semiconductors \cite{RevModPhys.55.645} and metals \cite{PhysRevB.99.205433} have been developed, but they also require the cell size and time step to be smaller than the mean free path and the relaxation time, and they are slow to converge and have a large statistical error at small Knudsen numbers; the lattice Boltzmann methods (LBM) \cite{guo_lattice_2013} have many applications to multiscale problems in nanosystems, but calculations have been shown to yield nonphysical solutions at high Knudsen numbers \cite{doi:10.1080/10407790.2014.915683}.

The finite-volume discrete unified gas kinetic scheme (DUGKS), which was originally developed for multiscale gas flows \cite{PhysRevE.88.033305,PhysRevE.91.033313}, is a multiscale method with asymptotic preservation (AP) properties, and has been developed to address multiscale transport problems of other energy carriers. Actually, the DUGKS has been successfully applied in various fields such as multiscale gas flow%\cite{zhu_discrete_2016,xin_discrete_2023}
, phonon heat transfer%\cite{guo_discrete_2016,luo_discrete_2017,zhang_discrete_2019}
, radiative heat transfer%\cite{PhysRevE.97.063302,song_discrete_2020}
, and plasma transport \cite{guo_progress_DUGKS}
%\cite{PhysRevE.101.043307}
. By coupling particle drift and scattering processes simultaneously in the flux reconstruction, this scheme allows the cell size and time step to be independent of the mean free path and relaxation time, and to recover adaptively from the ballistic limit to the diffusive limit \cite{PhysRevE.88.033305,PhysRevE.91.033313}. 
Very recently, some progress has been made in electron-phonon coupled heat transfer based on DUGKS \cite{zhang2023electronphonon}. 
However, it cannot be used to describe thermoelectric transport. Also, the single relaxation model used fails to capture the effects of electron momentum-conserving scattering.

In this study, we extend the application of DUGKS to study electrical, thermal and thermoelectric transport of electrons, incorporating the electronic band structure to iteratively solve for the temperature and chemical potential distributions. Numerical results are carefully compared to theoretical and numerical results. The advantage of DUGKS in solving multiscale problems is illustrated. The rest of the paper is organized as follows: Section \ref{BTE} describes the electron Boltzmann equation, Section \ref{Methods} details the DUGKS for the electron Boltzmann equation, Section \ref{results} verifies the performance of the proposed DUGKS by simulating several typical problems, and conclusions are given in Section \ref{conclusion}.

%In this work, we generalize DUGKS to electron transport, and constructed an efficient and accurate multiscale scheme for observing mesoscopic electron transport phenomena. By calculating four test cases, cross-plane electron thermal transport, in-plane electron transport, thermoelectric transport, and hydrodynamics of DC and AC, and comparing them with analytical or numerical solutions, the accuracy of this scheme is verified. 

\section{Electron Boltzmann transport equation} \label{BTE}

The Boltzmann transport equation describes the evolution of the quasi-particle distribution driven by external fields applied to the system. The transport properties can be readily calculated once the distribution function is known. 
The eBTE is expressed as:
\begin{equation}
\frac{\partial f}{\partial t}+\boldsymbol{v}\cdot \nabla _{\boldsymbol{r}}f+\boldsymbol{\dot{k}}\cdot \nabla _{\boldsymbol{k}}f=\left( \frac{\partial f}{\partial t} \right) _{\rm coll}.
\end{equation}
The distribution function $f=f_n(\boldsymbol{r}, \boldsymbol{k}, t)$ depends on band index $n$, position $\boldsymbol{r}$, the wave vector $\boldsymbol{k}$ and time $t$. Here we write it in an equivalent form $f=f\left( \boldsymbol{r},\varepsilon_n ,\boldsymbol{s},t \right) $ where $\varepsilon_n$ represents energy of the $n$-th band (in the following, for convenience of expression, we contract the band index $n$, $\boldsymbol{s}$ represents the unit direction angle. For 3D case, $\boldsymbol{s}=\left( \cos \theta ,\sin \theta \cos \varphi ,\sin \theta \sin \varphi \right) $, where $\theta$ and $\varphi$ are the polar and azimuthal angle. In 2D case, only $\varphi$ is needed $\boldsymbol{s}=\left( \cos \varphi ,\sin \varphi \right) $.  The group velocity of electrons depends on the band structure and is given by $\boldsymbol{v}={\nabla _{\boldsymbol{k}}\varepsilon_n\left(\boldsymbol{k} \right)}/{\hbar }$. The electric field $\boldsymbol{E}$ determines the time derivative $\dot{\boldsymbol{k}}= -e\boldsymbol{E/}{\hbar }$. The right hand side presents the change of the distribution function due to electron collisions. It depends on the specific scattering mechanism, which can be expressed as
%
%The first term on the left side of the equation represents the evolution of the distribution function over time, 
%the second term represents the change in the distribution function due to the movement of electrons in real space, caused by the temperature difference and the chemical potential difference, where the group speed $\boldsymbol{\dot{r}}={\nabla _{\boldsymbol{k}}\varepsilon \left( \boldsymbol{k} \right)}/{\hbar }$;
%the third term represents the change in the distribution function due to the movement of electrons in the inverse space, usually caused by the applied electric and magnetic fields, $\boldsymbol{\dot{k}}$ is the electron velocity in inverted space, and the right end of the equation is the change in $f$ due to collisions,
%$\left( {\partial f}/{\partial t} \right) _{coll}$ dependent on the specific scattering mechanism, which can be expressed as
\begin{equation}
\left(\frac{\partial f}{\partial t}\right)_{{coll }}=-\int \frac{d \boldsymbol{k}^{\prime}}{(2 \pi)^{3}}\left\{W\left(\boldsymbol{r}, \boldsymbol{k}, \boldsymbol{k}^{\prime}\right) f(\boldsymbol{k}) \left[1-f\left(\boldsymbol{k}^{\prime}\right)\right]-W\left(\boldsymbol{k}^{\prime}, \boldsymbol{k}\right) f\left(\boldsymbol{k}^{\prime}\right) [1-f(\boldsymbol{k})]\right\} ,
\end{equation}
where $W\left(\boldsymbol{k}, \boldsymbol{k}^{\prime}\right)$ is the scattering kernel, which describes the rate of electron transition from state $\boldsymbol{k}$ to $\boldsymbol{k}^{\prime}$, and is usually obtained from the Fermi's golden rule.

%Since the energy band structure, distribution function, and scattering process of electrons are described by quantum mechanics and the equations of motion satisfy the classical Newtonian equations, eBTE is a semi-classical transport equation for the case where the material is large enough so that the quantum effect can be neglected, which treats electrons as identical particles, consider only two-body interactions and assume statistical independence between colliding entirely homogeneous particles.
%It can be seen that the collision term is complex, while the equation itself is a seven-dimensional partial differential equation containing multiple integrals and is nonlinear for the distribution function, which makes the numerical solution very complicated.
The Boltzmann equation is a complicated nonlinear integral-differential equation which is difficult to solve even numerically.  
In this paper, we adopt the simplified Callaway approximation to describe the collision term, which is divided into two types, representing normal (N) and Umklapp (U) processes, respectively. The former fulfills crystal momentum conservation, while the latter does not due to the involved extra reciprocal lattice vector. A rich set of transport phenomena can be obtained from this simplified approximation, including ballistic, diffusive and hydrodynamic regimes.

Generally, the electric potential $\varphi$ and the electric field $\boldsymbol{E}=-\nabla \varphi$ needs to be obtained by solving the Poisson's equation. However, when the external field is relatively weak, the electric field can be represented by the gradient of the electrochemical potential, i.e., $\nabla \mu \approx -e\nabla \varphi $. We use this approximation and avoid solving the Poisson equation \cite{PhysRevB.99.165409,PhysRevLett.126.076803,PhysRevLett.101.216807}. 
%We further limit ourselves to the case driven by external temperature or/and chemical potential gradient, so that we can set $\boldsymbol{\dot{k}}=0$, meanwhile, the electric field can be characterized in the linear region by the chemical potential gradient \cite{PhysRevB.99.165409,PhysRevLett.126.076803,PhysRevLett.101.216807}.
%Although the electron motion in the metal generates an induced electric field, the contribution of this component to electron transport is much smaller than the drive of electrons by temperature or chemical potential differences, so 
The resulting eBTE can be written as 
\begin{equation}
  \frac{\partial f}{\partial t}+\boldsymbol{v}\cdot \nabla _{\boldsymbol{r}}f=\frac{f_{0}^{U}\left( T^U,\mu ^U \right) -f}{\tau ^U}+\frac{f_{0}^{N}\left( T^N,\mu ^N,\boldsymbol{u} \right) -f}{\tau ^N}.
  \label{equ:1}
\end{equation}
Here, $\tau^U$ and $\tau^N$ are the relaxation times of U-process and N-process, respectively, and both depend on the electron state. The Fermi-Dirac distribution $f_{0}^{U}$ and the shifted Fermi-Dirac distribution $f_{0}^{N}$ with a common drift velocity $\boldsymbol{u}$ for all electrons are given by:

\begin{equation}
f_{0}^{U}=\frac{1}{\exp \left[ \left( \varepsilon -\mu ^U \right) /k_BT^U \right] +1},
\end{equation}
\begin{equation}
f_{0}^{N}=\frac{1}{\exp \left[ \left( \varepsilon -\mu ^N-\boldsymbol{p}\cdot \boldsymbol{u} \right) /k_BT^N \right] +1}.
\end{equation}

%$f_{0}^U=\left\{ \exp \left[ \left( \varepsilon -\mu^U \right) /k_BT^U \right] +1 \right\} ^{-1}$ is the Fermi-Dirac distribution and $f_{0}^{N}=\left\{ \exp \left[ \left( \varepsilon -\mu ^N-\boldsymbol{p}\cdot \boldsymbol{u} \right) /k_BT^N \right] +1 \right\} ^{-1}$ is the shifted Fermi-Dirac distribution with a common drift velocity $\boldsymbol{u}$ for all electrons.

%The scattering to which electrons are subjected in metals is mainly from phonon scattering, due to the large number of free electrons scattering frequently with the lattice, we first assume that the electron-phonon scattering is isotropic, 
%secondly assume that phonons are always in local equilibrium and fully thermalized with electrons, so the scattering term satisfies both particle number conservation and energy conservation

The two sets of parameters $\left\{ \mu ^U,T^U \right\}$ and $\left\{ \mu ^N,T^N,\boldsymbol{u} \right\}$ in $f_0^U$ and $f_0^N$ are Lagrange multipliers in the maximum entropy principle, responsible for the conservation of particle number and energy in U-processes and the conservation of particle number, energy and momentum in N-processes, respectively. They are determined by the following equations
%In addition to the scattering terms of the two processes satisfying the conservation of energy and particle number, respectively, the N-process also satisfies the conservation of momentum, as follows:
\begin{eqnarray}
\int_{4\pi}{\int_{-\infty}^{+\infty}{\boldsymbol{\psi }^U\frac{D\left( \varepsilon \right)}{4\pi}\frac{f_{0}^{U}\left( T^U,\mu ^U \right) -f}{\tau ^U\left( \varepsilon \right)}}}d\varepsilon d\Omega =\boldsymbol{0},
  \label{equ:3}
\end{eqnarray}

\begin{eqnarray}
\int_{4\pi}{\int_{-\infty}^{+\infty}{\boldsymbol{\psi }^N\frac{D\left( \varepsilon \right)}{4\pi}\frac{f_{0}^{N}\left( T^N,\mu ^N,\boldsymbol{u}^N \right) -f}{\tau ^N\left( \varepsilon \right)}}}d\varepsilon d\Omega =\boldsymbol{0}, 
  \label{equ:4}  
\end{eqnarray}
where $\boldsymbol{\psi }^U=\left( 1,\varepsilon \right) ^T$ and $\boldsymbol{\psi }^N=\left( 1,\varepsilon ,\boldsymbol{p} \right) ^T$ denote the conserved quantities of the U-process and N-process, respectively, $D\left( \varepsilon \right)$ is the electronic density of states, $d\Omega$ is the differential of the direction angle.
These equations can be used to get $\left\{ \mu ^U,T^U \right\}$ and $\left\{ \mu ^N,T^N,\boldsymbol{u} \right\}$,  once the nonequilibrium distribution function $f$ is obtained.
%$T_{\rm loc}$ and $\mu _{\rm loc}$ can be obtained by associating Eqs.~(\ref{equ:3}) and (\ref{equ:4}).

The temporal parameters $T^U$, $T^N$, $\mu^U$ and $\mu^N$ with the dimensions of temperature and chemical potential can be different from the physical ones. On the other hand,  the corresponding physical quantities $\tilde{T}$ and $\tilde{\mu}$ are determined from the local conservation laws,

\begin{equation}
  \int_{-\infty}^{+\infty}{\int_{4\pi}{\frac{D\left( \varepsilon \right)}{4\pi}f_0\left( \tilde{T}_{loc},\tilde{\mu} _{loc} \right) d\Omega d\varepsilon}}=\int_{-\infty}^{+\infty}{\int_{4\pi}{\frac{D\left( \varepsilon \right)}{4\pi}fd\Omega d\varepsilon}} ,
  \label{equ:20.1}
\end{equation}

\begin{equation}
  \int_{-\infty}^{+\infty}{\int_{4\pi}{\varepsilon \frac{D\left( \varepsilon \right)}{4\pi}f_0\left( \tilde{T}_{loc},\tilde{\mu}_{loc} \right) d\Omega d\varepsilon}}=\int_{-\infty}^{+\infty}{\int_{4\pi}{\varepsilon \frac{D\left( \varepsilon \right)}{4\pi}fd\Omega d\varepsilon}} .
  \label{equ:20.2}
\end{equation}
It can be found that the distribution function $f$ reduces to $f^U$ and $f^N$ in the diffusive and hydrodynamic regimes, respectively, while the intermediate regime it is a weighted average of the two. If $\tau$ is constant, $\tilde{T}$ and $\tilde{\mu}$ calculated in the diffusive and hydrodynamic regions are the same as the spurious variables.

\section{The discrete unified gas kinetic scheme} \label{Methods}
%\subsection{The discrete unified gas kinetic scheme}
We introduce the DUGKS for solving Eq.~(\ref{equ:1}) numerically. To facilitate the solution, we rewrite Eq.~(\ref{equ:1}) with discrete angular space as

\begin{equation}
\frac{\partial f {(}\boldsymbol{r,}\varepsilon ,\boldsymbol{s}_{\alpha},t {)}}{\partial t}+\boldsymbol{v}\cdot \nabla _{\boldsymbol{r}}f {(}\boldsymbol{r,}\varepsilon ,\boldsymbol{s}_{\alpha},t {)}=\frac{f_0-f {(}\boldsymbol{r,}\varepsilon ,\boldsymbol{s}_{\alpha},t {)}}{\tau} ,
\end{equation}
where $
f_0=\left( \tau ^Nf_{0}^{U}+\tau ^Uf_{0}^{N} \right) /\left( \tau ^N+\tau ^U \right) 
$ and $
\tau =\tau ^U\tau ^N/\left( \tau ^U+\tau ^N \right) 
$. In order to accurately evaluate the zeroth-order to second-order moments of the distribution function, the discrete angle $\boldsymbol{s}_{\alpha}$ needs to satisfy the following requirements:
\begin{equation}
\sum_{\alpha}{w_{\alpha}}=\Omega ,\ \ \ \ \sum_{\alpha}{w_{\alpha}\boldsymbol{s}_{\alpha}}=\boldsymbol{0},\ \ \ \ \sum_{\alpha}{w_{\alpha}\boldsymbol{s}_{\alpha}\boldsymbol{s}_{\alpha}}=\frac{\Omega}{d}\boldsymbol{I},
\end{equation}
where $w_{\alpha}$ is the weight of the angular discretization, $d$ is the system dimension. For $d=2$, we have $\Omega=2\pi $; for $d=3$, we have $\Omega=4\pi$, and $\boldsymbol{I}$ is the corresponding unit matrix.

Similar to the calculation of phonon Boltzmann transport equation, we use the trapezoidal integration rule to discretize the energy space. For the direction angle space, the conventional $S_N$ quadrature is not accurate enough for large Knudsen number and may have a serious ''ray effect", To overcome these difficulties, we choose the Gauss–Legendre (G-L) rule to discretize the direction angle.
The real space is discretized using the finite volume method, the mid-point rule is used for the time integration of the advection term, and the trapezoidal rule is used for the collision term. With these considerations, Eq.~(\ref{equ:1}) is discretized as
\begin{equation} \label{equ:2}
    f_{\alpha ,\varepsilon ,i}^{n+1}-f_{\alpha ,\varepsilon ,i}^{n}+\frac{\Delta t}{V_i}\boldsymbol{F}_{\alpha ,\varepsilon ,i}^{n+1/2}=\frac{\Delta t}{2}\left[ \frac{f_{0,\varepsilon ,i}^{n+1}-f_{\alpha ,\varepsilon ,i}^{n+1}}{\tau _{\varepsilon}}+\frac{f_{0,\varepsilon ,i}^{n}-f_{\alpha ,\varepsilon ,i}^{n}}{\tau _{\varepsilon}} \right] ,
\end{equation}
where $f_{\alpha ,\varepsilon ,i}^{n}$ denotes the cell-averaged occupation probability of electrons moving along the $\boldsymbol{s}_{\alpha}$ direction in the cell $i$ at the energy level $\varepsilon$ at time $t=n\Delta t$,
$V_i$ is the volume of cell $i$ and $\boldsymbol{F}_{\alpha ,\varepsilon ,i}^{n+1/2}$ is the flux across the interfaces of cell $i$, is expressed as
\begin{equation} 
    \boldsymbol{F}_{\alpha ,\varepsilon ,i}^{n+1/2}=\sum_{j\in \mathcal{N}_i}{\left( \boldsymbol{v}_{\alpha ,\varepsilon}\cdot \boldsymbol{n}_{ij} \right) f_{\alpha ,\varepsilon}^{n+1/2}\left( \boldsymbol{x}_{ij} \right) S_{ij}} ,
    \label{equ:13}
\end{equation}
where $\mathcal{N}_i$ denotes the set of cells adjacent to cell $i$, $\boldsymbol{n}_{ij}$ is the unit normal vector pointing from cell $i$ to cell $j$, $S_{ij}$ is the area of the interface $ij$ between cells $i$ and $j$, and $f^{n+1/2}\left( \boldsymbol{x}_{ij} \right)$ denotes the distribution function at the interface at the time of $t_{n+1/2}=t_n+\Delta t/2$.
Two new distribution functions are introduced to remove the implicitness of Eq.~(\ref{equ:2})
\begin{eqnarray}
    \tilde{f}_{\alpha ,\varepsilon ,i}^{n}=f_{\alpha ,\varepsilon ,i}^{n}-\frac{\Delta t}{2}\left( \frac{f_{0,\varepsilon ,i}^{n}-f_{\alpha ,\varepsilon ,i}^{n}}{\tau _{\varepsilon}} \right),
    \label{equ:15}
    \\
    \tilde{f}_{\alpha ,\varepsilon ,i}^{+,n}=f_{\alpha ,\varepsilon ,i}^{n}+\frac{\Delta t}{2}\left( \frac{f_{0,\varepsilon ,i}^{n}-f_{\alpha ,\varepsilon ,i}^{n}}{\tau _{\varepsilon}} \right) .
\end{eqnarray} 
Then, Eq. (\ref{equ:2}) can be rewritten as
\begin{equation}  \label{equ:5}
    \tilde{f}_{\alpha ,\varepsilon ,i}^{n+1}=\tilde{f}_{\alpha ,\varepsilon ,i}^{+,n}-\frac{\Delta t}{V_i}\boldsymbol{F}_{\alpha ,\varepsilon ,i}^{n+1/2} .
\end{equation}

We can track the evolution of the distribution function $\tilde{f}$ following Eq.~(\ref{equ:5}), where the interface distribution function $f_{ij}^{n+1/2}$ at half-time steps is reconstructed based on the eBTE, which is the key difference between the present DUGKS and classical DOM using certain direct numerical interpolations. First, along the characteristic line of Eq.~(\ref{equ:1}) from time $t_n$ to $t_{n+1/2}$, the end point $\boldsymbol{x}_{ij}$ at the center of the interface between cell $i$ and cell $j$.
\begin{equation}  \label{equ:6}
f_{\alpha ,\varepsilon}^{n+1/2}\left( \boldsymbol{x}_{ij} \right) -f_{\alpha ,\varepsilon}^{n}\left( \boldsymbol{x}_{ij}^{'} \right) =\frac{\Delta t}{4}\left[ \frac{f_{0,\varepsilon}^{n+1/2}\left( \boldsymbol{x}_{ij} \right) -f_{\alpha ,\varepsilon}^{n+1/2}\left( \boldsymbol{x}_{ij} \right)}{\tau _{\varepsilon}}+\frac{f_{0,\varepsilon}^{n}\left( \boldsymbol{x}_{ij}^{\prime} \right) -f_{\alpha ,\varepsilon}^{n}\left( \boldsymbol{x}_{ij}^{\prime}\right)}{\tau _{\varepsilon}} \right] ,
\end{equation}
where $\boldsymbol{x}_{ij}^{\prime}=\boldsymbol{x}_{ij}-\boldsymbol{v}\Delta t/2$. Again introducing two auxiliary distribution functions to remove the implicitness of Eq. (\ref{equ:6})
\begin{eqnarray}
    \bar{f}_{\alpha ,\varepsilon ,i}^{n}=f_{\alpha ,\varepsilon ,i}^{n}-\frac{\Delta t}{4}\left( \frac{f_{0,\varepsilon ,i}^{n}-f_{\alpha ,\varepsilon ,i}^{n}}{\tau _{\varepsilon}} \right),
    \label{equ:8}
    \\
    \bar{f}_{\alpha ,\varepsilon ,i}^{+,n}=f_{\alpha ,\varepsilon ,i}^{n}+\frac{\Delta t}{4}\left( \frac{f_{0,\varepsilon ,i}^{n}-f_{\alpha ,\varepsilon ,i}^{n}}{\tau _{\varepsilon}} \right) ,
    \label{equ:9}
\end{eqnarray}
it can be expressed as
\begin{equation}
    \bar{f}_{\alpha ,\varepsilon}^{n+1/2}\left( \boldsymbol{x}_{ij} \right) =\bar{f}_{\alpha ,\varepsilon}^{+,n}\left( \boldsymbol{x}_{ij}^{\prime} \right) .
    \label{equ:7}
\end{equation}

\begin{figure}
\centering
\includegraphics[width=0.42\textwidth]{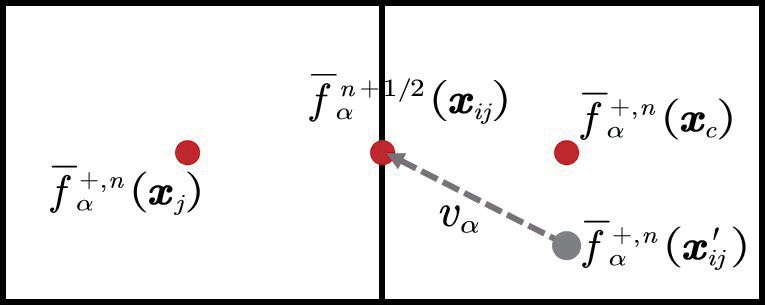}
\caption{\label{fig:pic} Schematic of reconstructed distribution function at the interface.}
\end{figure}

To evaluate the interface flux $\boldsymbol{F}_{\alpha ,\varepsilon ,i}^{n+1/2}$, we assume that the electron distribution function varies linearly in each cell to reconstruct the auxiliary function $\bar{f}_{\alpha ,\varepsilon}^{+,n}\left( \boldsymbol{x}_{ij}^{\prime} \right)$ in Eq.~(\ref{equ:7}), as shown in Fig.~\ref{fig:pic}, then

\begin{equation} 
    \bar{f}_{\alpha ,\varepsilon}^{+,n}\left( \boldsymbol{x}_{ij}^{\prime} \right) =\bar{f}_{\alpha ,\varepsilon}^{+,n}\left( \boldsymbol{x}_c \right) +\left( \boldsymbol{x}_{ij}^{\prime}-\boldsymbol{x}_c \right) \cdot \boldsymbol{\sigma }_c ,
\label{equ:10}
\end{equation}
where $\boldsymbol{\sigma }_c$ is the slope of the auxiliary function $\bar{f}_{\alpha ,\varepsilon}^{+,n}\left( \boldsymbol{x}_{ij}^{\prime} \right)$ in the cell where $\boldsymbol{x}_c$ is located. We have $c=i$ if $\boldsymbol{v}_{\alpha}\cdot \boldsymbol{n}_{ij}>0$ and $c=j$ otherwise, which can be constructed smoothly by methods such as central-difference or van Leer limiter.
In this work, we use the latter to construct the gradient to ensure numerical accuracy and stability. In 1D case, the van Leer limiter is defined as
\begin{equation}
    \boldsymbol{\sigma }_i=\left[ {\rm sgn} \left( s_1 \right) +{\rm sgn} \left( s_2 \right) \right] \frac{\left| s_1 \right|\left| s_2 \right|}{\left| s_1 \right|+\left| s_2 \right|} ,
    \label{equ:20}
\end{equation}
where
\begin{equation}
    s_1=\frac{\bar{f}_{\alpha ,\varepsilon ,i}^{+,n}-\bar{f}_{\alpha ,\varepsilon ,i-1}^{+,n}}{x_i-x_{i-1}},\ \ s_2=\frac{\bar{f}_{\alpha ,\varepsilon ,i+1}^{+,n}-\bar{f}_{\alpha ,\varepsilon ,i}^{+,n}}{x_{i+1}-x_i} .
\end{equation}

As a result, the original distribution function ${f}_{\alpha ,\varepsilon}^{n+1/2}\left( \boldsymbol{x}_{ij} \right) $ can be obtained from Eqs.~(\ref{equ:8}), (\ref{equ:7}) and (\ref{equ:10})
\begin{equation}
    f_{\alpha ,\varepsilon}^{n+1/2}\left( \boldsymbol{x}_{ij} \right) =\frac{4\tau}{4\tau +\Delta t}\bar{f}_{\alpha ,\varepsilon}^{n+1/2}\left( \boldsymbol{x}_{ij} \right) +\frac{\Delta t}{4\tau +\Delta t}f_{0,\alpha ,\varepsilon}^{n+1/2}\left( \boldsymbol{x}_{ij} \right) ,
    \label{equ:14}
\end{equation}
where $f_{0,\alpha ,\varepsilon}^{n+1/2}\left( \boldsymbol{x}_{ij} \right)$ is a function of $\mu _{loc,ij}^{n+1/2}$ and $T_{loc,ij}^{n+1/2}$, which can be obtained by the particle number and energy conservation of the scattering operator. 
Since the original distribution function $f$ in Eqs.~(\ref{equ:3}) and (\ref{equ:4}) is unknown, the conservation equation needs to be converted to the form of $\bar{f}$ from Eqs.~(\ref{equ:3}), (\ref{equ:4}) and (\ref{equ:8})
\begin{eqnarray}
\int_{4\pi}{\int_{-\infty}^{+\infty}{\boldsymbol{\psi }^U\frac{D\left( \varepsilon \right)}{4\pi}\left[ \frac{\left( 4\tau ^N+\Delta t \right) f_{0,ij}^{U,n+1/2}-\Delta tf_{0,ij}^{N,n+1/2}}{\left( 4\tau +\Delta t \right) \left( \tau ^U+\tau ^N \right)}-\frac{4\tau \bar{f}_{ij}^{n+1/2}}{\left( 4\tau +\Delta t \right) \tau ^U} \right]}}d\varepsilon d\Omega =\boldsymbol{0} ,
    \label{equ:11}
\end{eqnarray}

\begin{eqnarray}
\int_{4\pi}{\int_{-\infty}^{+\infty}{\boldsymbol{\psi }^N\frac{D\left( \varepsilon \right)}{4\pi}\left[ \frac{\left( 4\tau ^N+\Delta t \right) f_{0,ij}^{N,n+1/2}-\Delta tf_{0,ij}^{U,n+1/2}}{\left( 4\tau +\Delta t \right) \left( \tau ^U+\tau ^N \right)}-\frac{4\tau \bar{f}_{ij}^{n+1/2}}{\left( 4\tau +\Delta t \right) \tau ^N} \right]}}d\varepsilon d\Omega =\boldsymbol{0} .
  \label{equ:12}  
\end{eqnarray}

If multiple bands contribute to transport, the above equation should also include a summation over band indices.
In this way, the interface flux $\boldsymbol{F}_{\alpha ,\varepsilon ,i}^{n+1/2}$ is fully determined by Eqs.~(\ref{equ:13}) and (\ref{equ:14}).
From Eq.~(\ref{equ:5}) we can obtain the distribution function $\tilde{f}_{\alpha ,\varepsilon ,i}^{n+1}$ at $t_{n+1}$. Similarly, $T_{loc,i}^{n+1}$ and $\mu _{loc,i}^{n+1}$ can be determined from the following two equations 
\begin{eqnarray}
\int_{4\pi}{\int_{-\infty}^{+\infty}{\boldsymbol{\psi }^U\frac{D\left( \varepsilon \right)}{4\pi}\left[ \frac{\left( 2\tau ^N+\Delta t \right) f_{0,i}^{U,n+1}-\Delta tf_{0,i}^{N,n+1}}{\left( 2\tau +\Delta t \right) \left( \tau ^U+\tau ^N \right)}-\frac{2\tau \tilde{f}_{i}^{n+1}}{\left( 2\tau +\Delta t \right) \tau ^U} \right]}}d\varepsilon d\Omega =\boldsymbol{0} ,
    \label{equ:16}
\end{eqnarray}

\begin{eqnarray}
\int_{4\pi}{\int_{-\infty}^{+\infty}{\boldsymbol{\psi }^N\frac{D\left( \varepsilon \right)}{4\pi}\left[ \frac{\left( 2\tau ^N+\Delta t \right) f_{0,i}^{N,n+1}-\Delta tf_{0,i}^{U,n+1}}{\left( 2\tau +\Delta t \right) \left( \tau ^U+\tau ^N \right)}-\frac{2\tau \tilde{f}_{i}^{n+1}}{\left( 2\tau +\Delta t \right) \tau ^N} \right]}}d\varepsilon d\Omega =\boldsymbol{0} .
  \label{equ:17}  
\end{eqnarray}

To solve Eqs.~\ref{equ:11}-\ref{equ:17}, we use the Newtonian iteration method.
The electric and heat currents in the center of each cell can be expressed as
\begin{equation}
  \boldsymbol{J}\left( \boldsymbol{x}_i,t \right) =-e\int_{4\pi}{\int_{\varepsilon _{\min}}^{\varepsilon _{\max}}{\boldsymbol{v}\frac{D\left( \varepsilon \right)}{4\pi}}\frac{2\tau \left( \varepsilon \right)}{2\tau \left( \varepsilon \right) +\Delta t}\tilde{f}\left( \boldsymbol{x}_i,t \right) d\varepsilon d\Omega} ,
  \label{equ:18.1}
\end{equation}
\begin{equation}
    \boldsymbol{J}_q\left( \boldsymbol{x}_i,t \right) =\int_{4\pi}{\int_{\varepsilon _{\min}}^{\varepsilon _{\max}}{\left[ \varepsilon -\tilde{\mu} \left( \boldsymbol{x}_i,t \right) \right] \boldsymbol{v}\frac{D\left( \varepsilon \right)}{4\pi}}\frac{2\tau \left( \varepsilon \right)}{2\tau \left( \varepsilon \right) +\Delta t}\tilde{f}\left( \boldsymbol{x}_i,t \right) d\varepsilon d\Omega} .
    \label{equ:18}
\end{equation}
%Then, we can define the effective thermal conductivity of the material: $\kappa _{eff}=\left| \boldsymbol{J}_q \right|L/\Delta T$，$L$ is the length of material. 
%
It is notable that the distribution functions $\tilde{f}^+$, $\bar{f}^+$ and $\tilde{f}$ satisfy the following relationship, which simplifies the calculation
\begin{equation}
    \tilde{f}^+=\frac{4}{3}\bar{f}^+-\frac{1}{3}\tilde{f} .
    \label{equ:19}
\end{equation}
Importantly, in DUGKS, the time step $\Delta t$ is determined by the Courant-Friedrichs-Lewy (CFL) condition
\begin{equation}
    \Delta t=\gamma \min \left( \frac{\Delta x}{\boldsymbol{v}} \right) ,
    \label{equ:25}
\end{equation}
where $\gamma \in \left( 0,1 \right)$ is the CFL number, such that the defined time step remains consistent for any relaxation time and has the AP property.

The procedure of the present DUGKS can be s summarized as follows:
\begin{enumerate}
    \item Set the initial temperature $T_0$, chemical potential $\mu_0$, discrete energy space, real space and direction angle space, and set the initial distribution function $\tilde{f}_{\alpha ,\varepsilon ,i}^{n}$ according to Eq.~(\ref{equ:15});
    \item Calculate $\bar{f}_{\alpha ,\varepsilon ,i}^{+,n}$ and its slope $\boldsymbol{\sigma }_c$, and construct the distribution function $\bar{f}_{\alpha ,\varepsilon}^{+,n}\left( \boldsymbol{x}_{ij}^{\prime} \right)$ according to Eq.~(\ref{equ:10});
    \item Calculate the interface distribution $\bar{f}_{\alpha ,\varepsilon}^{n+1/2}\left( \boldsymbol{x}_{ij} \right)$ based on Eq.~(\ref{equ:20});
    \item Calculate the local temperature $T_{loc,ij}^{n+1/2}$ and local chemical potential $\mu _{loc,ij}^{n+1/2}$ at the cell interface based on Eqs. (\ref{equ:11}) and (\ref{equ:12}) to obtain the corresponding equilibrium distribution $f_{0,\alpha ,\varepsilon}^{n+1/2}\left( \boldsymbol{x}_{ij} \right)$;
    \item Calculate the original distribution function $f_{0,\alpha ,\varepsilon}^{n+1/2}\left( \boldsymbol{x}_{ij} \right)$ at the cell interface based on Eq.~(\ref{equ:14}), and update the cell interface flux $\boldsymbol{F}_{\alpha ,\varepsilon ,i}^{n+1/2}$ by Eq. (\ref{equ:13});
    \item Calculate $\tilde{f}_{\alpha ,\varepsilon ,i}^{+,n}$ based on Eq.~(\ref{equ:19}) to update $\tilde{f}_{\alpha ,\varepsilon ,i}^{n+1}$ at the new time step by Eq.~(\ref{equ:5});
    \item Update the temperature $T_{loc,i}^{n+1}$ and chemical potential $\mu_{loc,i}^{n+1}$ of the cell at the next time step based on Eqs.~(\ref{equ:16}) and (\ref{equ:17});
    \item 
    %If needed, the equivalent parameters $\tilde{T}$ and $\tilde{\mu}$ can be calculated first by Eqs.~(\ref{equ:20.1}) and (\ref{equ:20.2}), and the current and heat flux can be calculated later by Eqs.~(\ref{equ:18}) and (\ref{equ:18.1}); otherwise 
    Repeat step 2 to step 8 until stop criterion is reached.
\end{enumerate}

\subsection{Boundary conditions}

For isothermal boundary conditions, electrons colliding with the boundary are absorbed, while an electron in equilibrium at the boundary temperature $T_b$ and chemical potential $\mu _b$ is emitted into the computational domain,
\begin{equation}
    f\left( \boldsymbol{x}_b,\varepsilon ,\boldsymbol{s} \right) =f_0\left( T_b,\mu _b,\varepsilon \right) ,\ \ \ \boldsymbol{s}\cdot \boldsymbol{n}_b>0
    \label{equ:26}
\end{equation}
where $\boldsymbol{n}_b>0$ is the unit normal vector pointing from the interface to the domain.

For periodic boundary conditions, driven by temperature and/or chemical potential gradient, an electron leaves one boundary while another electron with the same velocity and energy enters the domain from the corresponding periodic boundary. The distribution of these two electrons deviates equally from the equilibrium state at the passing boundary,
\begin{equation}
    f\left( \boldsymbol{x}_{b1},\varepsilon ,\boldsymbol{s} \right) -f_0\left( T_{b1},\mu _{b1},\varepsilon \right) =f\left( \boldsymbol{x}_{b2},\varepsilon ,\boldsymbol{s} \right) -f_0\left( T_{b2},\mu _{b2},\varepsilon \right) 
    \label{equ:27}
\end{equation}
where $b1$ and $b2$ denote the corresponding periodic boundaries, respectively.

For diffusive reflection boundary, it is assumed that electrons reflected from the boundary are isotropic
\begin{equation}
    f\left( \boldsymbol{x}_b,\varepsilon ,\boldsymbol{s} \right) =\frac{\int_{\boldsymbol{s}^{\prime}\cdot \boldsymbol{n}_b<0}{\left( \boldsymbol{s}^{\prime}\cdot \boldsymbol{n}_b \right)}f\left( \boldsymbol{x}_b,\varepsilon ,\boldsymbol{s}^{\prime} \right) d\Omega}{\int_{\boldsymbol{s}^{\prime}\cdot \boldsymbol{n}_b>0}{\boldsymbol{s}^{\prime}\cdot \boldsymbol{n}_b}d\Omega},\ \ \ \boldsymbol{s}\cdot \boldsymbol{n}_b>0 .
    \label{equ:28}
\end{equation}
Meanwhile, for specular reflection boundary, we have
\begin{equation}
  f\left( \boldsymbol{x}_b,\varepsilon ,\boldsymbol{s} \right) =f\left( \boldsymbol{x}_b,\varepsilon ,\boldsymbol{s}^{''} \right) ,\ \ \ \boldsymbol{s}\cdot \boldsymbol{n}_b>0
  \label{equ:28.1}
\end{equation}
where $\boldsymbol{s}^{''}=\boldsymbol{s}-2\left( \boldsymbol{s}\cdot \boldsymbol{n}_b \right) \boldsymbol{n}_b$.
It is noted that both diffusive and specular boundaries no energy and particle exchanges occur at the interface, and are therefore also called adiabatic boundaries.

\section{Numerical results} \label{results}

In this section, numerical calculation of four types of problems are performed to test the performance of DUGKS. 
For the first two examples, we consider the cross-plane (Subsec.~\ref{subsec:cross}) and in-plane (Subsec.~\ref{subsec:inplane}) electron transport of the Au films to verify the DUGKS solution against the deviational MC scheme available in the literature \cite{PhysRevB.99.205433}.
In the third example (Subsec.~\ref{subsec:hydro}), we consider hydrodynamic transport in 2D systems using parameters of graphene.  
In the last example (Subsec.~\ref{subsec:thermoelectric}), we consider quasi-1D thermoelectric transport in model metal and semiconductor systems. 

In all calculations, we take into account electrons in the energy window of $\left[ \mu_+ +15k_BT_0,\mu_- -15k_BT_0 \right]$, with $\mu_+$ and $\mu_-$ the higher and lower chemical potential, respectively. The energy window is uniformly discretized. The grid number is denoted as $N_\varepsilon$. 
In the direction angle space, $\theta$ and $\varphi$ are discretized into $N_{\theta}$ and $N_{\varphi}$ subdirections using G-L rule. 

Note that if real physical quantities are used in the computation, large round off errors may occur due to the large disparity of their magnitudes. Therefore, dimensionless quantities are employed in our computation, where the following non-dimensional variables are employed, 
$$
\varepsilon ^*=\frac{\varepsilon}{\varepsilon _F},\ \ \ \mu ^*=\frac{\mu}{\varepsilon _F},\ \ \ T^*=\frac{T}{T_0},\ \ \ \boldsymbol{p}^*=\frac{\boldsymbol{p}}{{\hbar }k_F},\ \ \ \boldsymbol{u}^*=\frac{{\hbar }k_F}{\varepsilon _F}\boldsymbol{u} ,
$$
$$
\boldsymbol{r}^*=\frac{\boldsymbol{r}}{L_{ref}},\ \ t^*=\frac{v_Ft}{L_{ref}},\ \ \boldsymbol{v}^*=\frac{\boldsymbol{v}}{v_F},\ \ \  {\rm Kn}^U=\frac{v_F\tau ^U}{L_{ref}},\ \ \  {\rm Kn}^N=\frac{v_F\tau ^N}{L_{ref}} ,
$$
where $*$ denotes that the quantity is dimensionless (omitted in the text for ease of presentation), Kn$^U$nd Kn$^N$ are the Knussen numbers of the U-process and the N-process, $\varepsilon _F$, $k_F$, and $v_F$ are the Fermi energy, Fermi wavevector, and Fermi velocity, respectively, and $L_{ref}$ is the characteristic length.

\subsection{Cross-plane electron thermal transport}
\label{subsec:cross}
The reciprocal lattice of Au is of bcc type, and the first Brillouin zone is a truncated octahedron. The Fermi surface, although distorted along the $\left< 111 \right>$ direction, is approximately spherical, so we use the nearly-free electron model $\varepsilon ={\hbar }^2\boldsymbol{k}^2/\left( 2m \right)$ and set the Fermi energy $\varepsilon_F$=5.51 eV. To compare with the analytical results, we focus only on the U-processes and set $\tau ^U=27.7\sqrt{\varepsilon /\varepsilon _F}$ fs, $\tau ^N\rightarrow \infty $. 
The CFL number is fixed at 0.7. 
The average temperature $T_0$ is set to $300$K. 

%\begin{figure}
%  \centering
%  \subfloat{\label{fig:error_N}\includegraphics[width=0.38\textwidth]{error_N.eps}}~~
%  \subfloat{\label{fig:error_a}\includegraphics[width=0.38\textwidth]{error_a.eps}}~~
%  \caption{Calculation error of effective thermal conductivity: (a) effects of real-space discretization; (b) effects of angle discretization.}
%\end{figure}

\begin{figure}
  \centering
  \subfloat{\label{fig:2a}\includegraphics[width=0.39\textwidth]{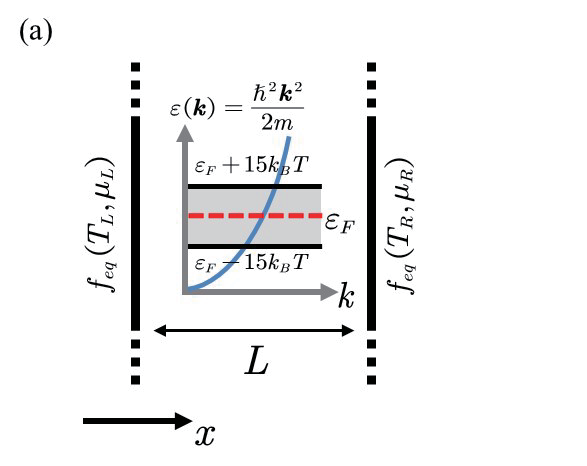}}~~
  \subfloat{\label{fig:2b}\includegraphics[width=0.37\textwidth]{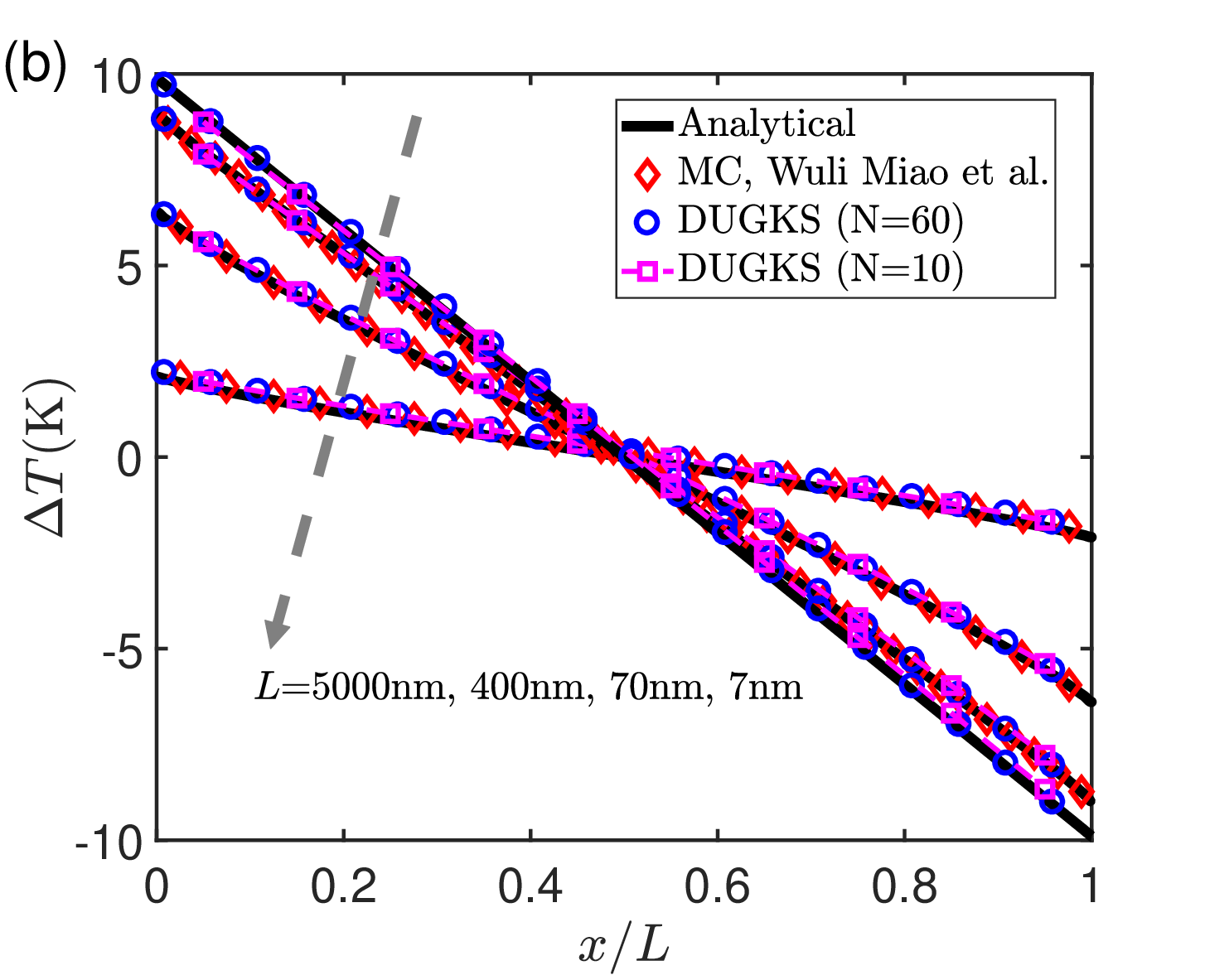}}~~\\
  \subfloat{\label{fig:2c}\includegraphics[width=0.37\textwidth]{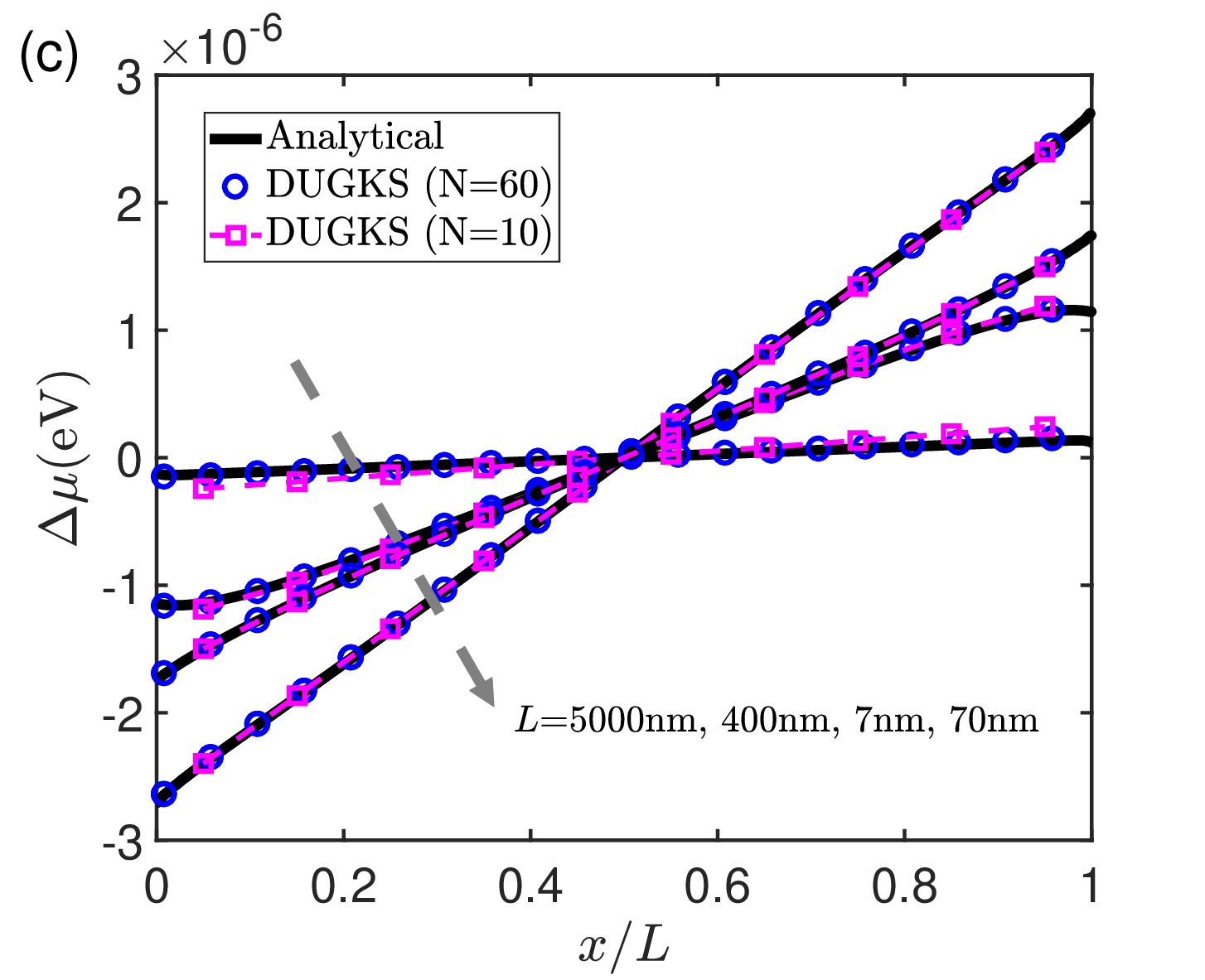}}~~
   \subfloat{\label{fig:2d}\includegraphics[width=0.39\textwidth]{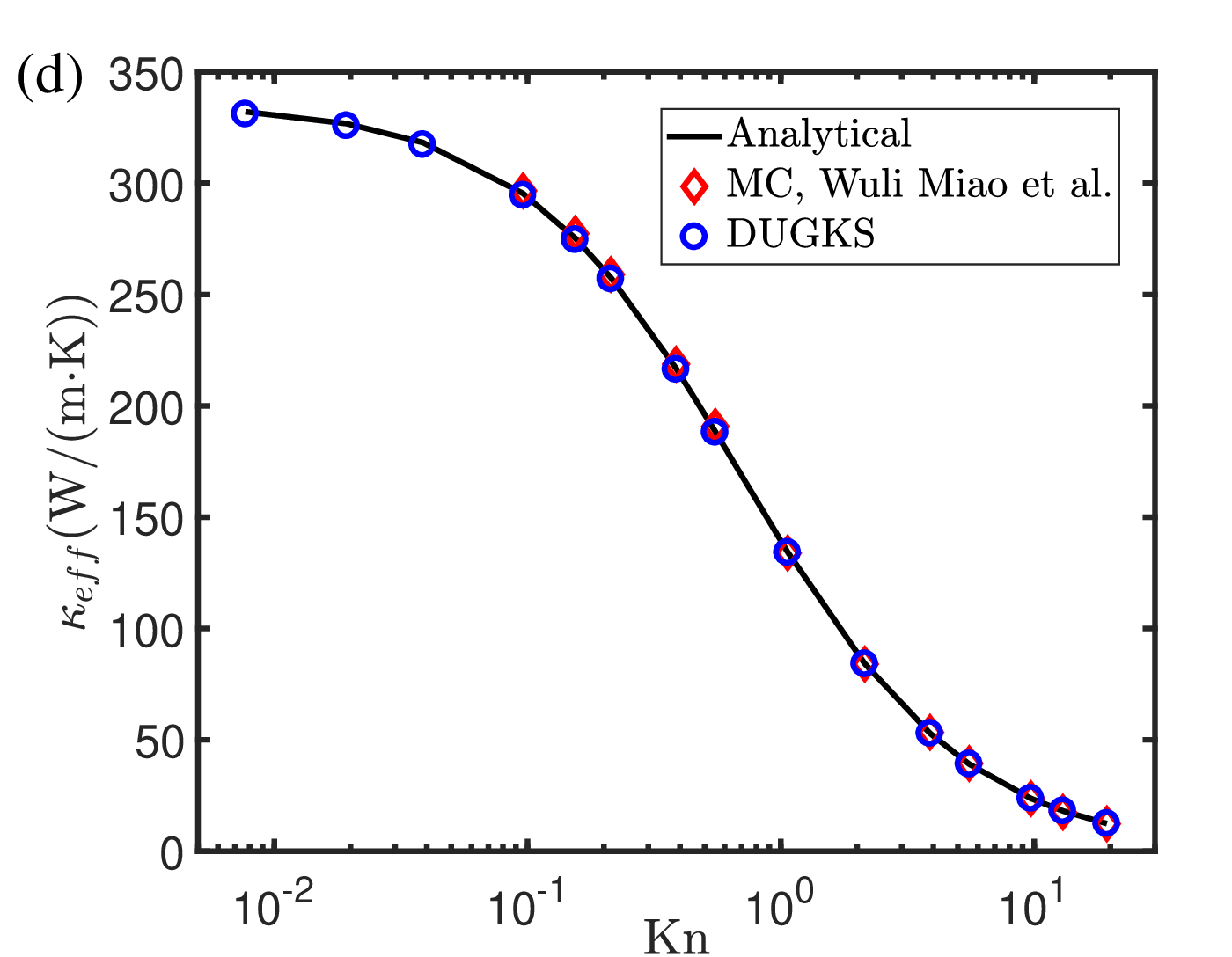}}~~
  \caption{
  Cross-plane electron thermal transport in Au films. 
  (a) Schematic of the system setup.  The energy band structure is shown in the middle. The gray area is the energy range considered in the calculation. We choose $\Delta T_L=-\Delta T_R=10$ K without initial chemical potential difference. (b) steady-state temperature distribution; (c) steady-state chemical potential distribution; (d) variation of effective thermal conductivity with Knudsen number.}
\end{figure}

We first consider the cross-plane ($x$ direction) electron thermal transport. The structure is shown in Fig.~\ref{fig:2a}. This is a quasi-1D problem since the system is translational invariant in $y$ and $z$ directions. 
We use isothermal boundary condition with $T_L=T_0+\Delta T_L$, $T_R=T_0+\Delta T_R$.
When $\Delta T_L$ and $\Delta T_R$ are relatively small, the distribution function can be linearized as $f\approx C_T\Delta T+C_{\mu}\Delta \mu $, with $C_T={\partial f_0}/{\partial T}$ and $C_{\mu}={\partial f_0}/{\partial \mu}$. A semi-analytical solution can then be obtained using the linear approximation, following Ref.~\cite{hua_semi-analytical_2015},
\begin{equation}
2\int_{\varepsilon _{\min}}^{\varepsilon _{\max}}{\frac{D\left( \varepsilon \right)}{\tau \left( \varepsilon \right)}\left( \begin{matrix}
	\varepsilon C_T \ \ \varepsilon C_{\mu}\\
	C_T \ \ C_{\mu}\\
\end{matrix} \right) d\varepsilon}\cdot \left( \begin{array}{c}
	\Delta T\left( \hat{x} \right)\\
	\Delta \mu \left( \hat{x} \right)\\
\end{array} \right) =F\left( \hat{x} \right) +\int_0^1{K\left( \hat{x},\hat{x}^{\prime} \right) \left( \begin{array}{c}
	\Delta T\left( \hat{x} \right)\\
	\Delta \mu \left( \hat{x} \right)\\
\end{array} \right) d\hat{x}^{\prime}} .
    \label{equ:29}
\end{equation}
This is the second type of Fredholm integral equation, with $\hat{x}=x/L$ the dimensionless coordinate, and
\begin{equation}
    F\left( \hat{x} \right) =\int_{\varepsilon _{\min}}^{\varepsilon _{\max}}{\frac{D\left( \varepsilon \right)}{\tau \left( \varepsilon \right)}\left( \begin{matrix}
        \varepsilon C_T\ \		\varepsilon C_{\mu}\\
        C_T\ \		C_{\mu}\\
    \end{matrix} \right) \left[ E_2\left( \frac{\hat{x}}{Kn_{\varepsilon}} \right) \left( \begin{array}{c}
        \Delta T_L\\
        \Delta \mu _L\\
    \end{array} \right) +E_2\left( \frac{1-\hat{x}}{Kn_{\varepsilon}} \right) \left( \begin{array}{c}
        \Delta T_R\\
        \Delta \mu _R\\
    \end{array} \right) \right] d\varepsilon},
    \label{equ:31}
\end{equation}

\begin{equation}
    K\left( \hat{x},\hat{x}^{\prime} \right) =\int_{\varepsilon _{\min}}^{\varepsilon _{\max}}{\frac{D\left( \varepsilon \right)}{Kn_{\varepsilon}\tau \left( \varepsilon \right)}E_1\left( \frac{\left| \hat{x}-\hat{x}^{\prime} \right|}{Kn_{\varepsilon}} \right) \left( \begin{matrix}
        \varepsilon C_T\ \		\varepsilon C_{\mu}\\
        C_T\ \		C_{\mu}\\
    \end{matrix} \right) d\varepsilon}.
    \label{equ:33}
\end{equation}
Here, $E_n\left( x \right) =\int_0^1{\theta ^{n-2}\exp \left( -x/\theta \right) d\theta}$, and $Kn_{\varepsilon}=v_{\varepsilon}\tau _{\varepsilon}/L$ is the energy-dependent Knudsen number. 
From the above equations, we can obtain $\Delta T(\hat{x})$ and $\Delta \mu(\hat{x})$ using the degenerate kernels method.

We can then calculate the steady-state heat flux density as
\begin{equation}
J_q=\int{\int_0^1{\left( \varepsilon -\varepsilon _{\text{F}} \right) f^+v\theta}d\theta d\varepsilon}-\int{\int_0^1{\left( \varepsilon -\varepsilon _{\text{F}} \right) f^-v\theta}d\theta d\varepsilon},
  \label{equ:Jq}
\end{equation}
where $f^+$ and $f^-$ describe the forward and backward transport of electrons, respectively, and can be written as
\begin{equation}
  f^+\left( \hat{x} \right) =\left( C_T\Delta T_L+C_{\mu}\Delta \mu _L \right) e^{-\frac{\hat{x}}{Kn_{\varepsilon}\theta}}+\int_0^{\hat{x}}{\frac{C_T\Delta T\left( \hat{x} \right) +C_{\mu}\Delta \mu \left( \hat{x} \right)}{Kn_{\varepsilon}\theta}e^{\frac{\hat{x}^{\prime} -\hat{x}}{Kn_{\varepsilon}\theta}}d\hat{x}^{\prime} },
  \label{equ:f+}
\end{equation}
\begin{equation}
  f^-\left( \hat{x} \right) =\left( C_T\Delta T_R+C_{\mu}\Delta \mu _R \right) e^{-\frac{1-\hat{x}}{Kn_{\varepsilon}\theta}}+\int_{\hat{x}}^1{\frac{C_T\Delta T\left( \hat{x} \right) +C_{\mu}\Delta \mu \left( \hat{x} \right)}{Kn_{\varepsilon}\theta}e^{\frac{\hat{x}^{\prime} -\hat{x}}{Kn_{\varepsilon}\theta}}d\hat{x}^{\prime} }.
  \label{equ:f-}
\end{equation}
The effective thermal conductivity of the system is then obtained from 
\begin{equation}
\kappa _{eff} = LJ_q/ \left| T_L-T_R \right| .   
\end{equation}

  \begin{figure}[t]
  \centering
  \subfloat{\label{fig:3a}\includegraphics[width=0.35\textwidth]{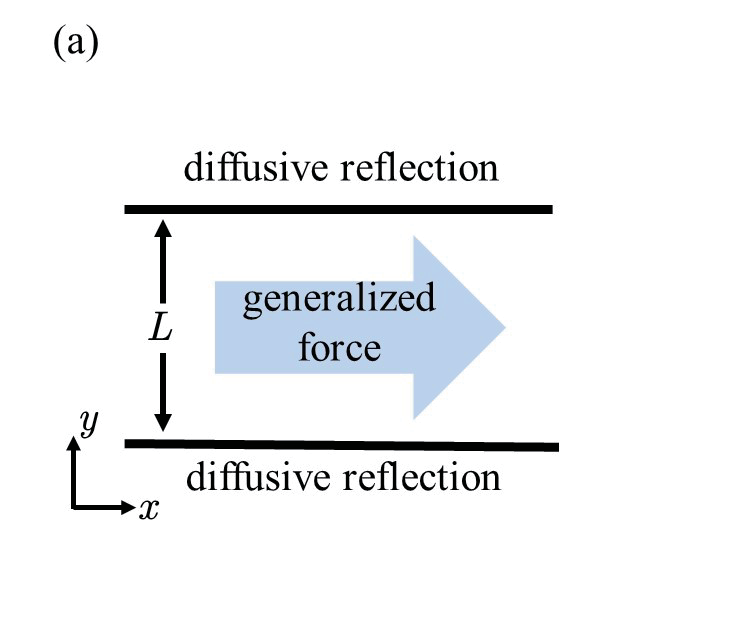}}~~
  \subfloat{\label{fig:3b}\includegraphics[width=0.38\textwidth]{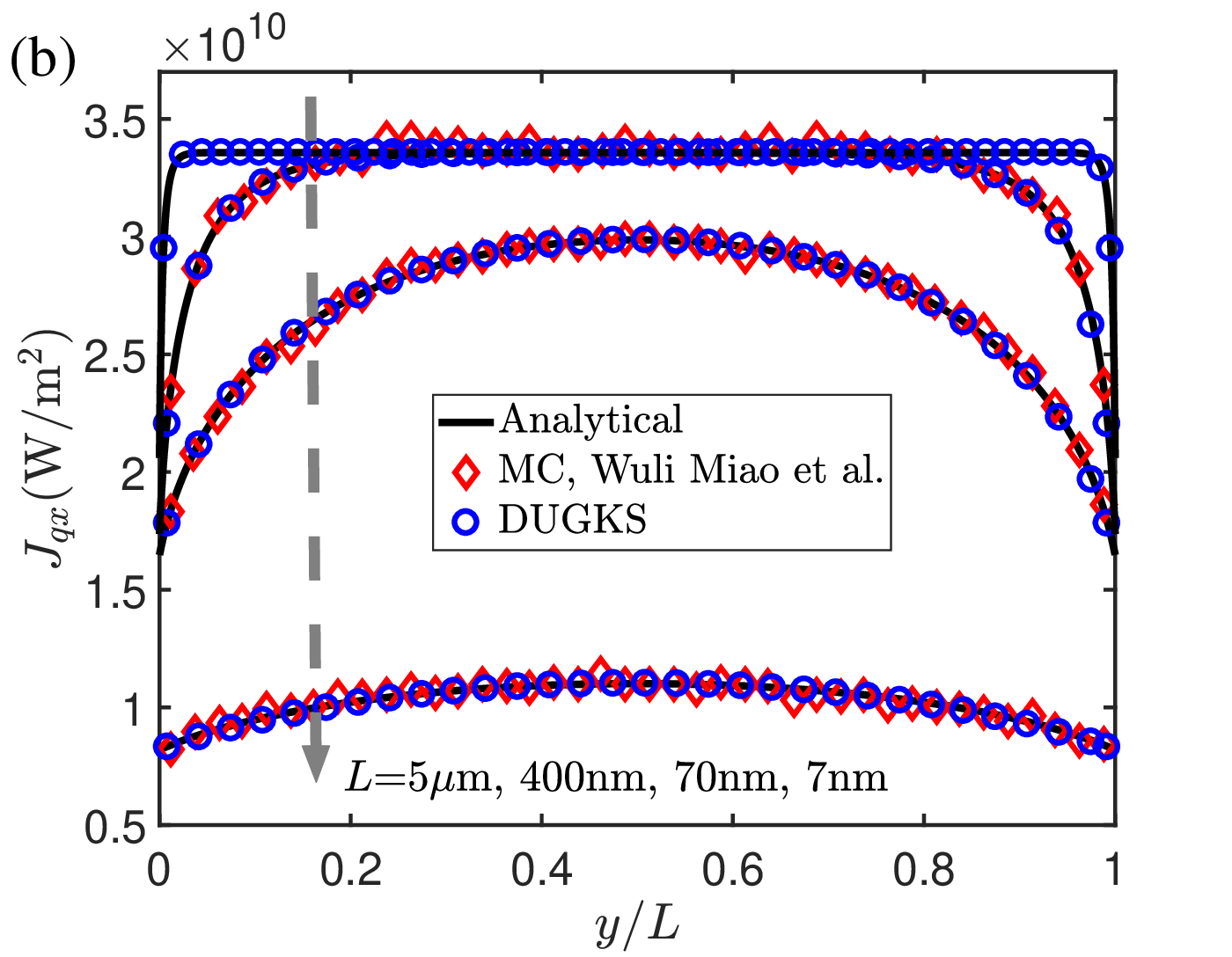}}~~\\
  \subfloat{\label{fig:3c}\includegraphics[width=0.38\textwidth]{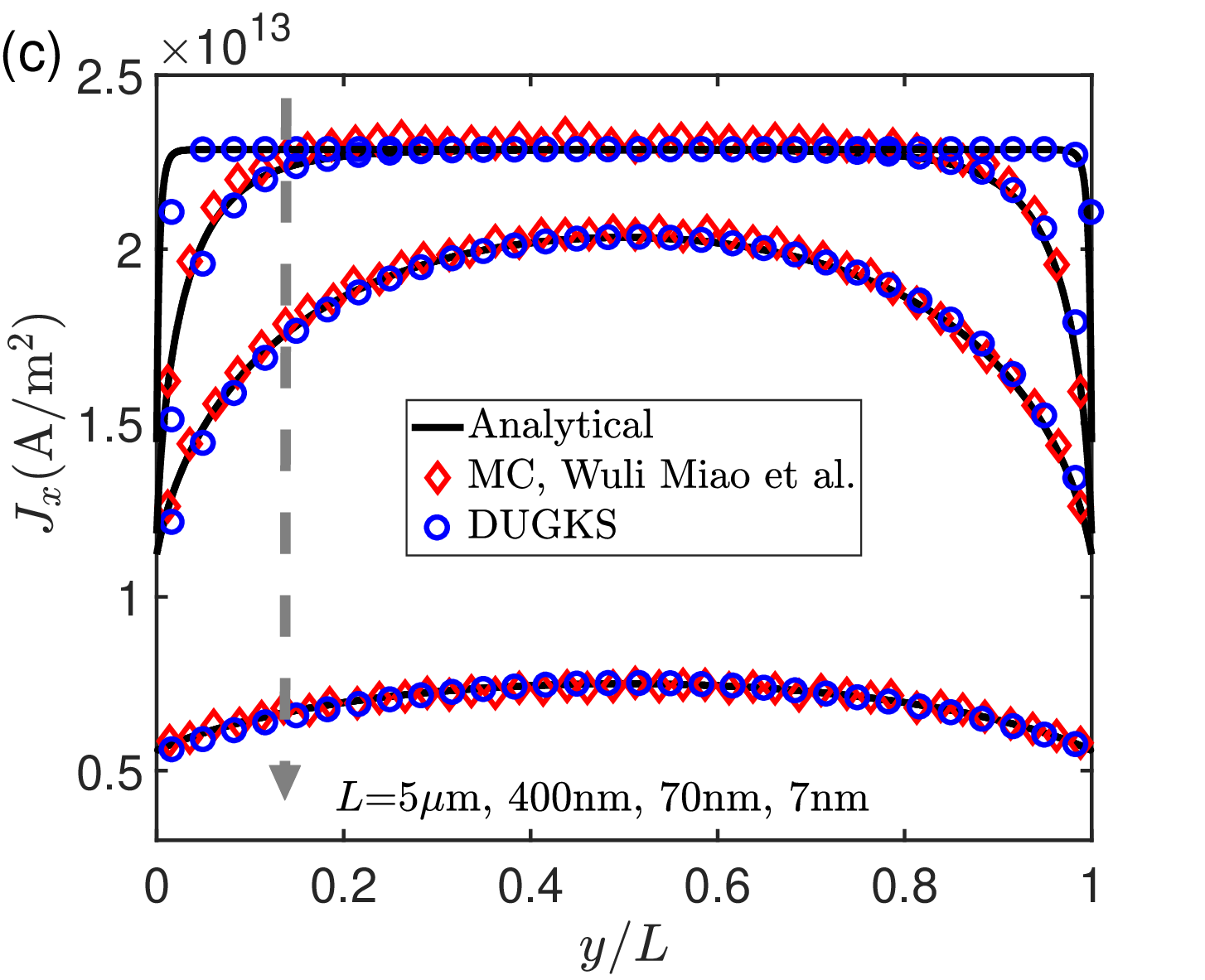}}~~
  \subfloat{\label{fig:3d}\includegraphics[width=0.38\textwidth]{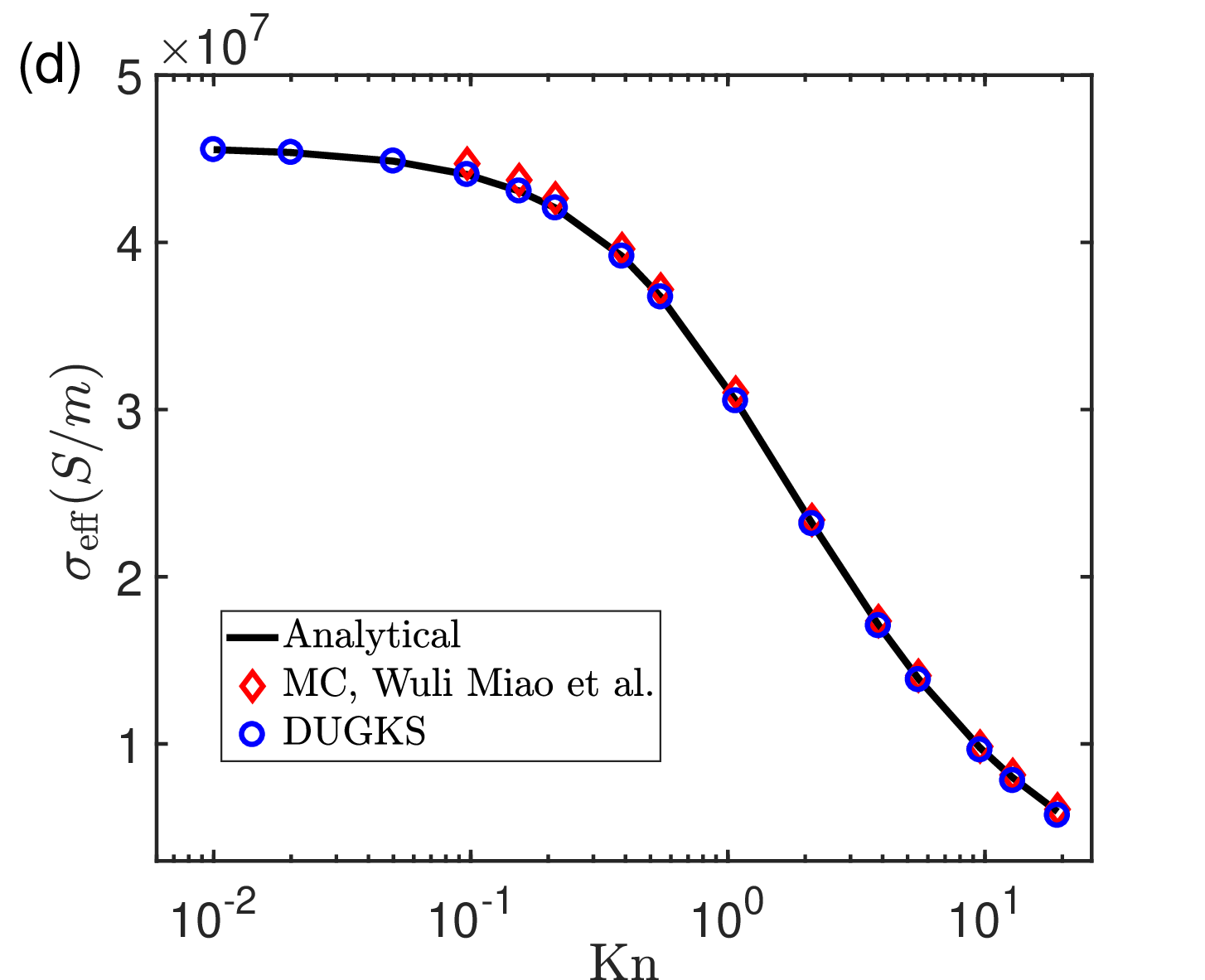}}~~
  \caption{In-plane electrical and thermal transport in Au films, with $L=5$ $\mu$m, 400 nm, 70 nm, 7 nm, covering both diffusive and ballistic transport regimes. The real space grids are set to $N_x\times N_y=2\times 60$, and the direction angle grids are $N_{\theta}\times N_{\varphi}=40\times 40$. (a) Schematic of the transport setup. (b) Heat flux density ($J_{qx}$) distribution along $y$ driven by temperature gradient in $x$ direction. (c) Current density ($J_x$) distribution along $y$ driven by a chemical potential gradient in $x$ direction. (d) The effective conductivity $\sigma _{\rm eff}=e\bar{J}_x/\left( d\mu /dx \right) $ as a function of Knudsen number. Here, $\bar{J}_x$ is the average of $J_x$ along $y$.}
\end{figure}

Different transport regimes can be characterized by 
the Knudsen number Kn $={\lambda}/{L_{\rm ref}}$, where $\lambda $ is the electron mean free path.
When Kn $\ll 1$, electrons are frequently scattered and exhibit diffusive transport; when Kn $\gg 1$, they move through the material almost without scattering and exhibit ballistic transport.
We consider several different lengths $L=5$ $\mu$m, 400 nm, 70 nm, 7 nm. The corresponding Kn is 0.0077, 0.096, 0.55, 5.5, respectively, going from diffusive to ballistic regime. We compare our numerical results with the above semi-analytical solution and deviational MC results \cite{PhysRevB.99.205433}. 
The direction angle is discretized into $N_{\theta}=100$ sub-directions using the G-L rule, and the system in $x$ direction is uniformly discretized to $N$ = 60 and 10, respectively.   
 The results are shown in Figs.~\ref{fig:2b}-\ref{fig:2d}. 
Fig.~\ref{fig:2b} shows that as the film size decreases, the system deviates from Fourier's law, and the boundary temperature slip becomes significant. As the mean free path of electrons increases, the electrons moving to the boundary are not sufficiently thermalized and strongly scatter with the electrons emitted from the boundary, giving rise to a nonlinear temperature distribution. Our scheme also captures the temperature-induced change in chemical potential (Fig.~\ref{fig:2c}), caused by the thermoelectric effect. For small temperature difference, this change is so tiny that we can ignore it when considering electron thermal transport. However, this small change is an essential factor in maintaining particle number conservation. The behavior of the effective thermal conductivity (Fig.~\ref{fig:2d}) is similar to that of the phonon case, with enhanced boundary scattering suppressing the thermal conductivity as the film thickness decreases.

Thus, we find good agreement of our numerical results  with the analytical solutions in the whole range from ballistic to diffusive transport, with good convergence on multiscale. 
Due to the coupled treatment of drift and scattering, in our scheme the mesh size does not have to be smaller than the particle mean free path. The results converge well even for sparse meshes with $N = 10$. In addition, the time step $\Delta t$ is completely determined by the CFL number and is not constrained by the relaxation time.
%, which in the case of $L=5$ $\mu$m is $\Delta t/\tau ^U$ = 1.46, and 8.78 for $N = $60 and 10, respectively.  
These illustrate that the scheme has AP property.

\begin{figure}
  \centering
  \subfloat{\label{fig:7a}\includegraphics[width=0.48\textwidth]{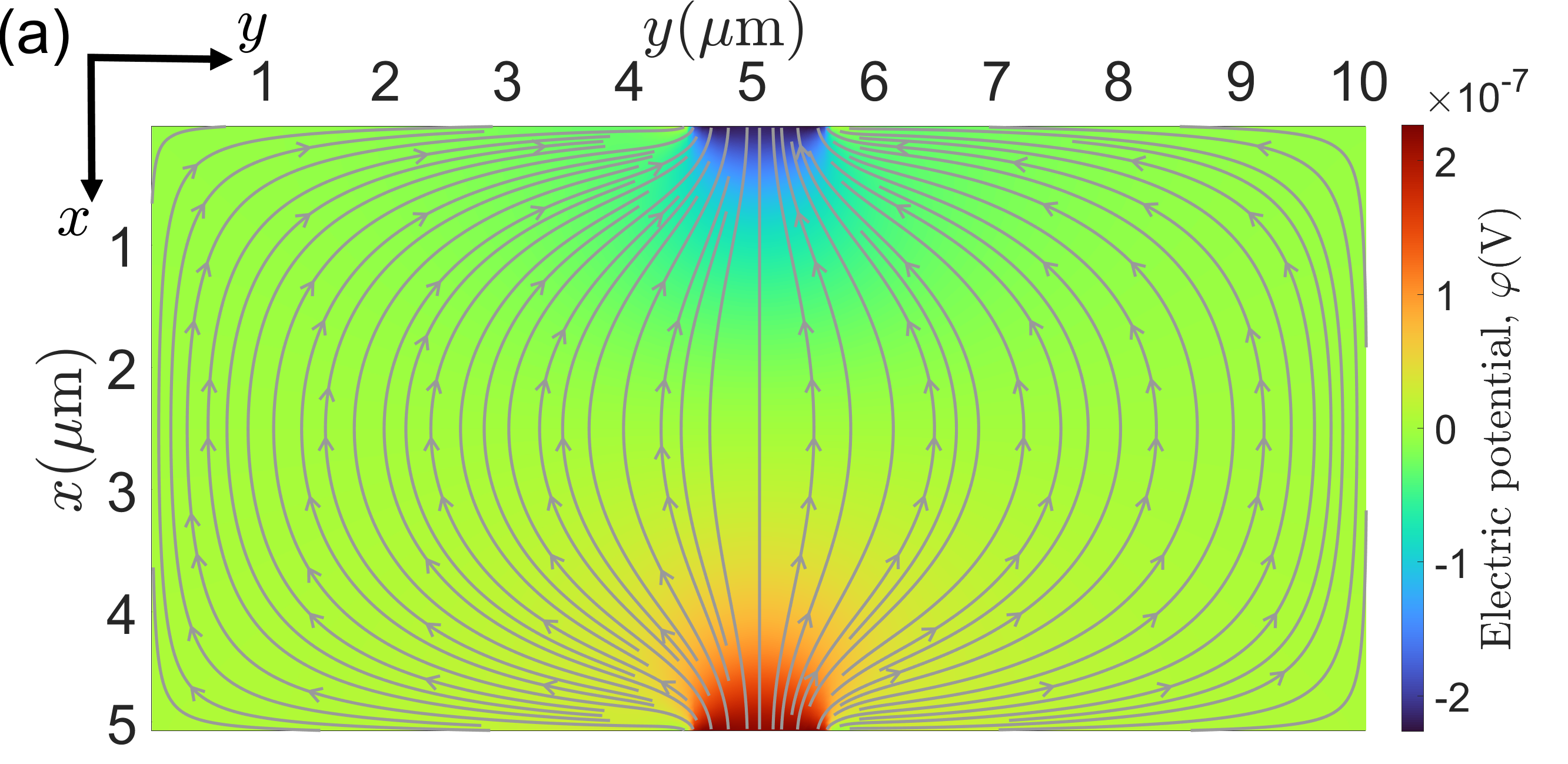}}~~
  \subfloat{\label{fig:7b}\includegraphics[width=0.50\textwidth]{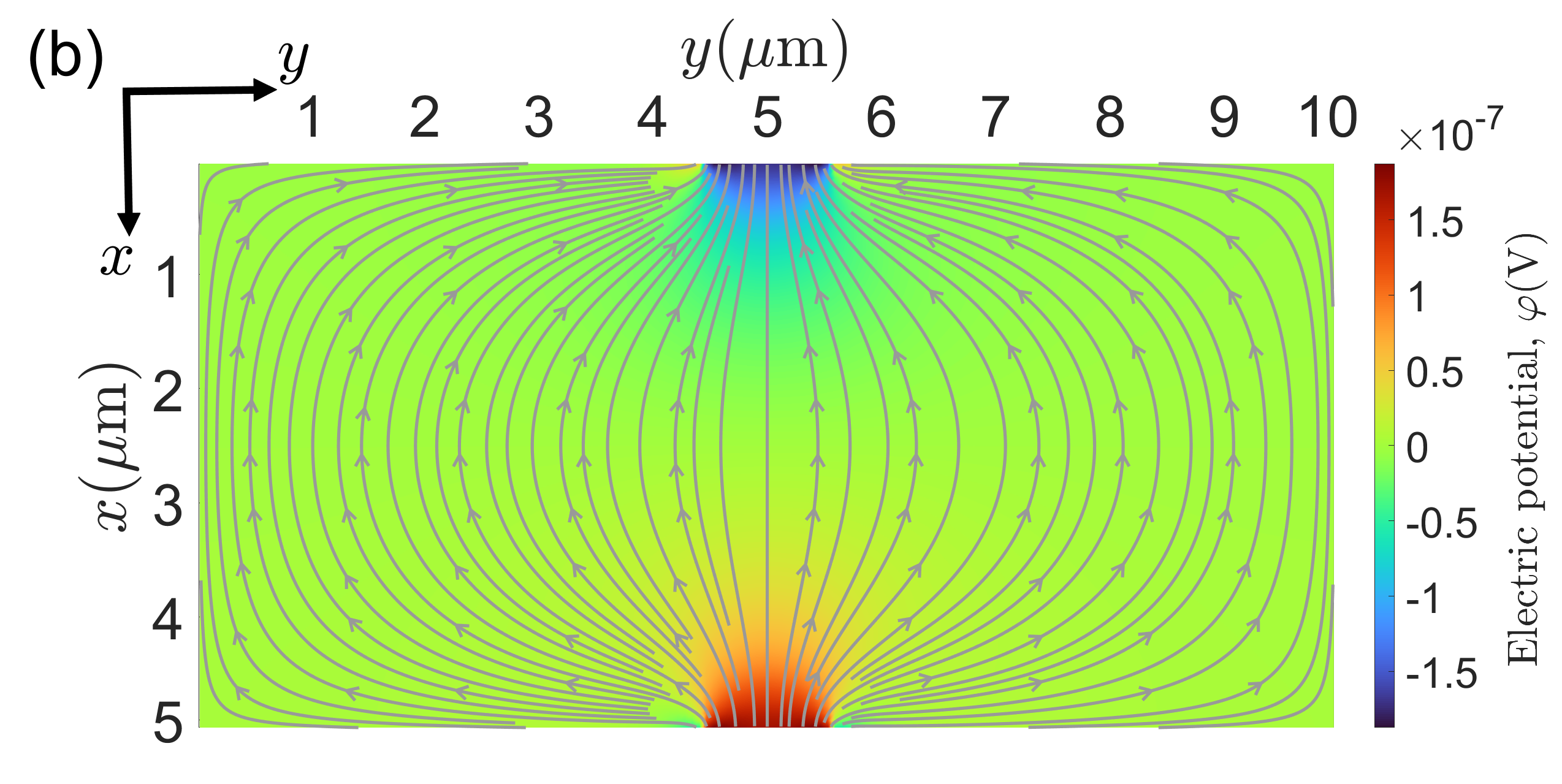}}~~\\
  \subfloat{\label{fig:7c}\includegraphics[width=0.52\textwidth]{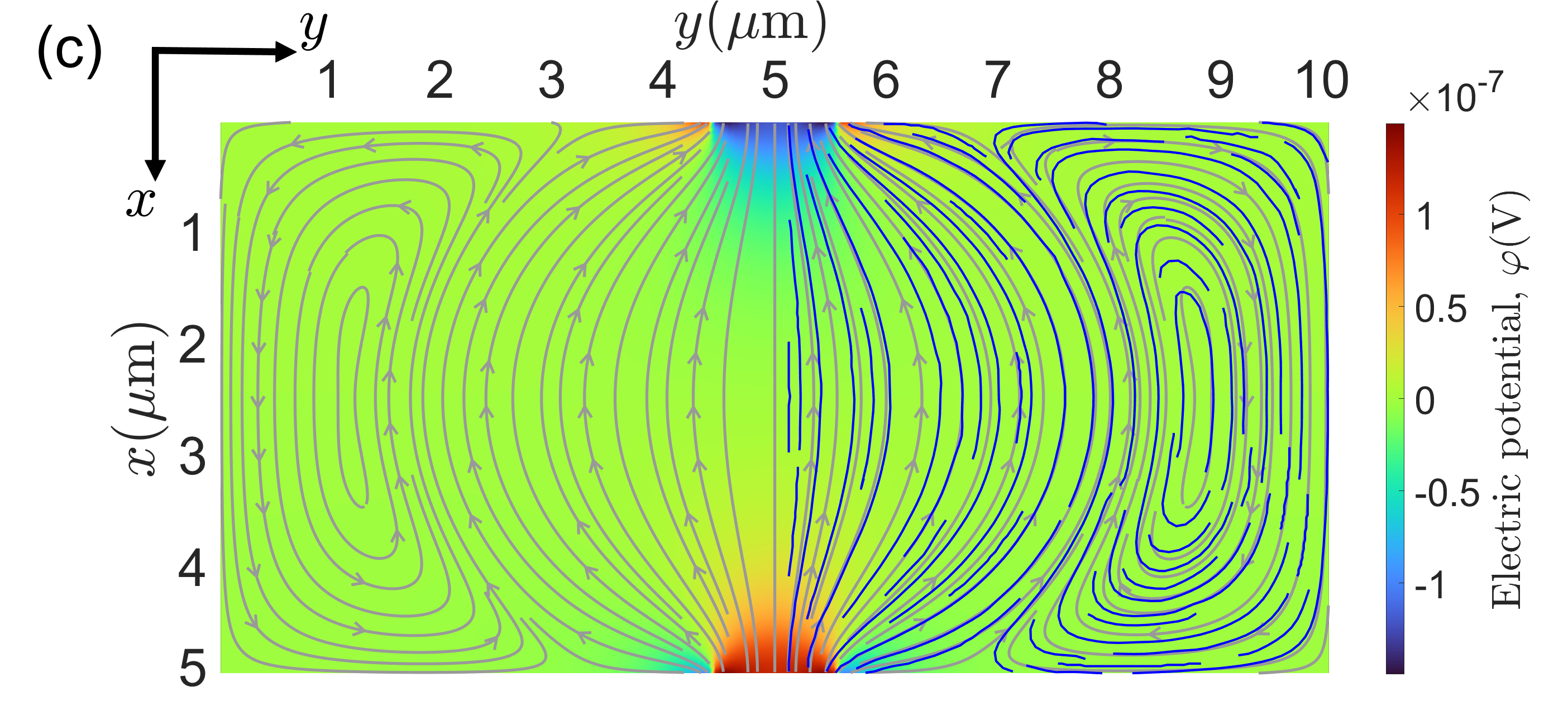}}~~
  \subfloat{\label{fig:7d}\includegraphics[width=0.48\textwidth]{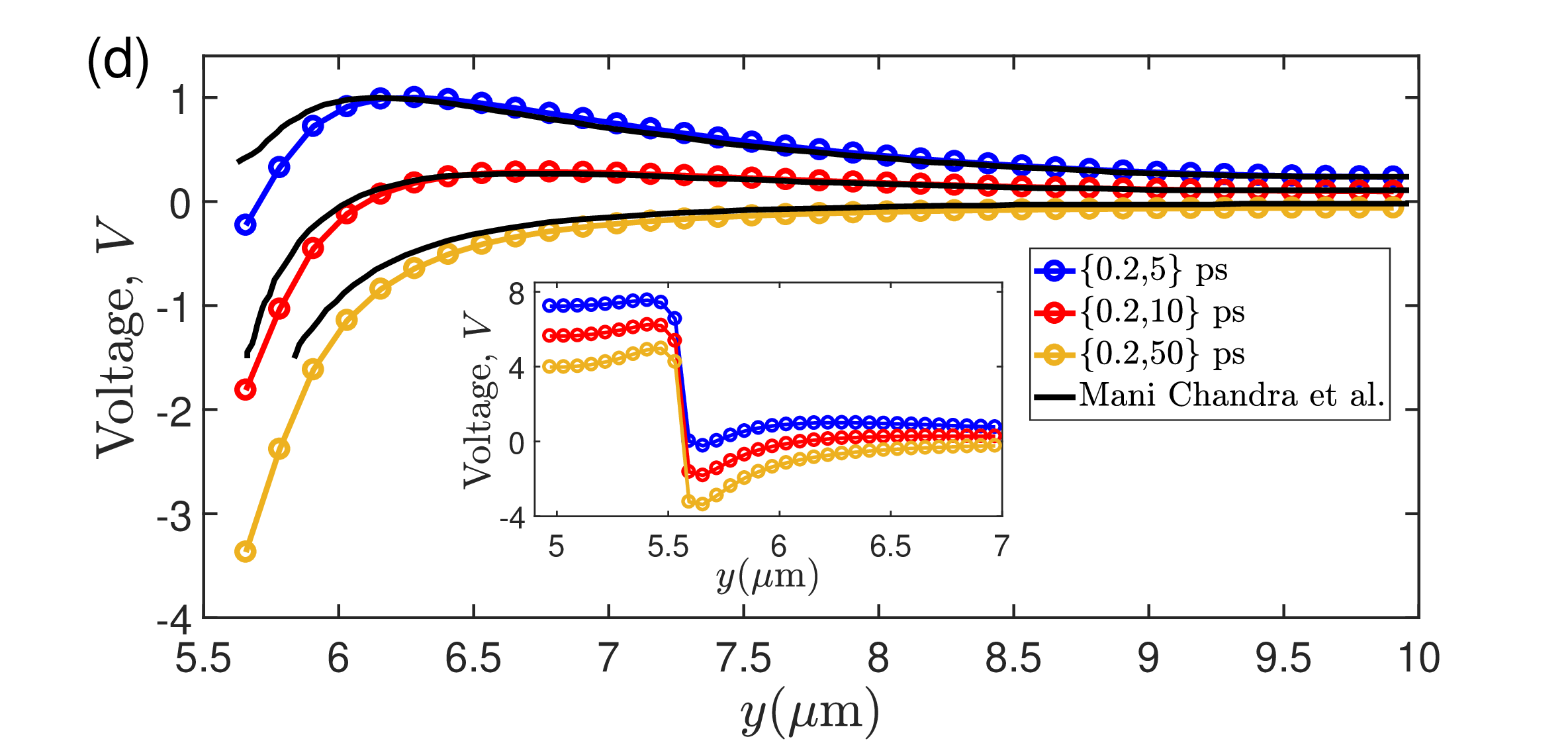}}~~
  \caption{\label{fig:7} DC and AC hydrodynamic transport in 2D sheet of size $5\times 10 \mu$m. 
  We use $N_{\varepsilon}=80$, $N_x\times N_y=80\times 160$ for energy and real space, respectively. For 2D angle space, we use $N_{\varphi}=400$. Potential distribution and current lines for $\left\{\tau^N,\tau^U\right\}$= (a) $\left\{ 0.2, 5 \right\}$ ps, (b) $\left\{ 0.2, 10 \right\}$ ps, (c) $\left\{ 0.2, 50 \right\}$ ps,  and the darker line in (c) is the result of Mani Chandra \emph{et al.}. (d) The potential difference between lower and upper boundary at $y\in \left[ 5.7,10 \right]$ $\mu$m. We scaled results for other cases by setting the maximum value from parameter set $\left\{ 0.2, 5 \right\}$ ps to 1. The subplot shows the potential distribution at $y\in \left[ 5.5,7 \right]$ $\mu$m, where the jump in value can be observed.}
\end{figure}

\subsection{In-plane electron transport}
\label{subsec:inplane}

In this subsection, as an example of quasi-2D transport, we calculate the in-plane thermal and electrical transport properties in Au thin films. The system setup is shown in Fig.~\ref{fig:3a}. We use the diffusive boundaries for the upper and lower sides, and periodic boundaries for the left and right sides. Due to  the applied generalized forces (e.g., temperature or chemical potential gradient), the distribution function deviates from the equilibrium one and is written as $f = f_0 + f_1$, with $f_0$ the equilibrium part, and $f_1$ a small deviation. It can be described by the Fuchs-Sondheimer theory \cite{fuchs_1938,doi:10.1080/00018735200101151}:
\begin{equation}
f_1=\left\{
\begin{aligned}
	-v_y\tau \left( \frac{\partial f_0}{\partial T}\frac{dT}{dy}+\frac{\partial f_0}{\partial \mu}\frac{d\mu}{dy} \right) \left[ 1-\exp \left( -\frac{x}{\tau v_x} \right) \right], \ \		v_x>0\\
-v_y\tau \left( \frac{\partial f_0}{\partial T}\frac{dT}{dy}+\frac{\partial f_0}{\partial \mu}\frac{d\mu}{dy} \right) \left[ 1-\exp \left( \frac{L-x}{\tau v_x} \right) \right],\ \		v_x<0
\end{aligned}
\right.
\end{equation}
The electrical and heat flux density can be calculated from $f_1$. 

We consider the temperature and chemical potential driven cases separately. The results for temperature-driven case with $dT/dx=-0.1$ K/nm are shown in Fig.~\ref{fig:3b}. This scheme again accurately captures the heat transfer process at different Knudsen numbers. In contrast to MC, the DUGKS results are free from random errors, and the mean free path does not limit the mesh size. The heat flux density saturates at $L=5$ $\mu$m, corresponding to diffusive transport, following Fourier's law. 
The suppression of heat flux at the boundaries is due to inelastic scattering occurring at the diffusive reflection boundary. The heat flux density distribution becomes uniform again in the ballistic regime. For chemical-potential-driven case, with $d\mu /dx=-e\nabla \varphi$, $-\nabla \varphi=5\times 10^5$ V/m, the results are shown in Figs.~\ref{fig:3c}-\ref{fig:3d}.  The agreement with analytical solutions again demonstrates the accuracy of this multiscale scheme. 

\subsection{DC and AC hydrodynamic transport}
\label{subsec:hydro}

\begin{figure}
  \centering
  \subfloat{\label{fig:8a}\includegraphics[width=0.48\textwidth]{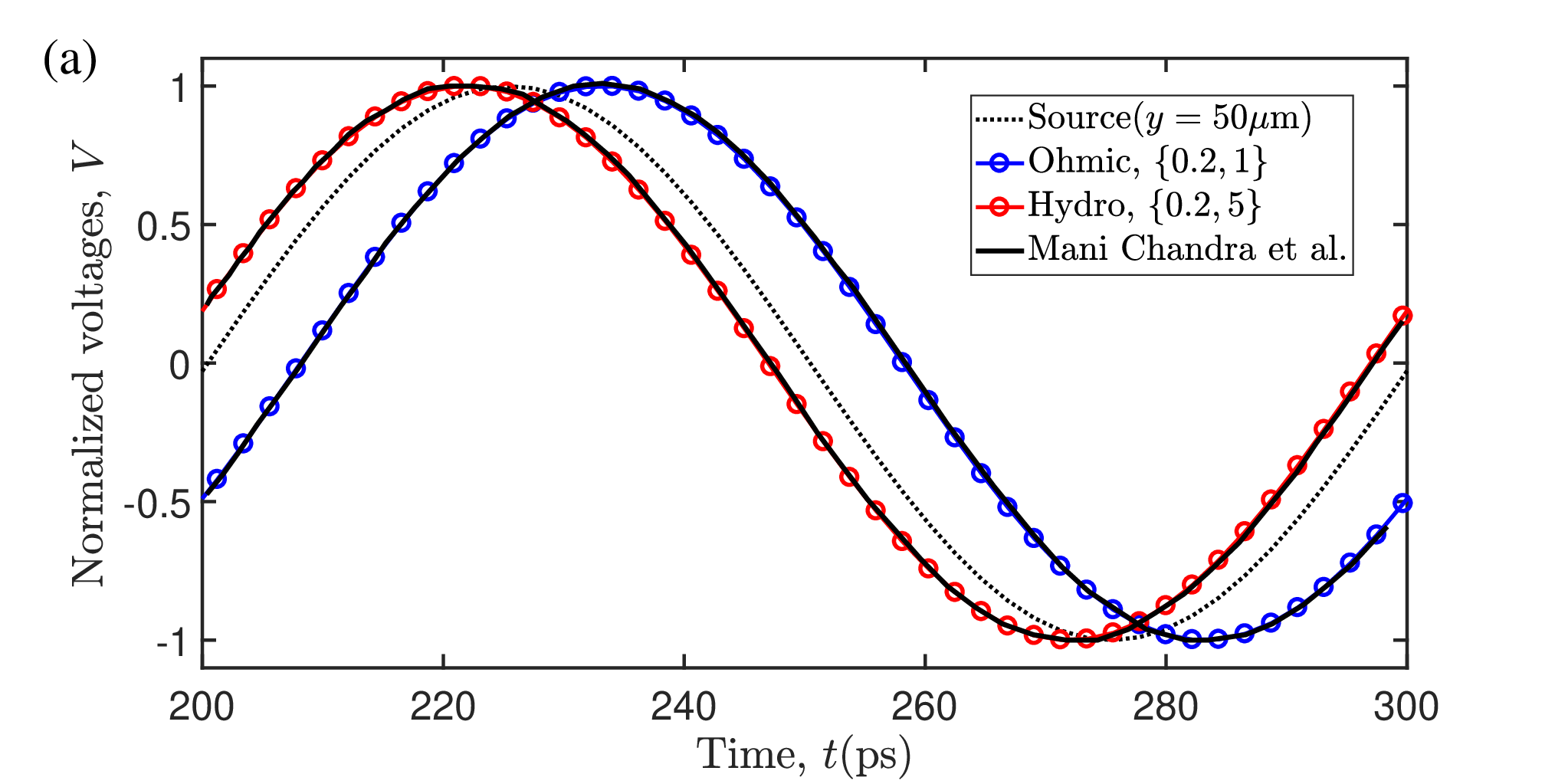}}~~
  \subfloat{\label{fig:8b}\includegraphics[width=0.48\textwidth]{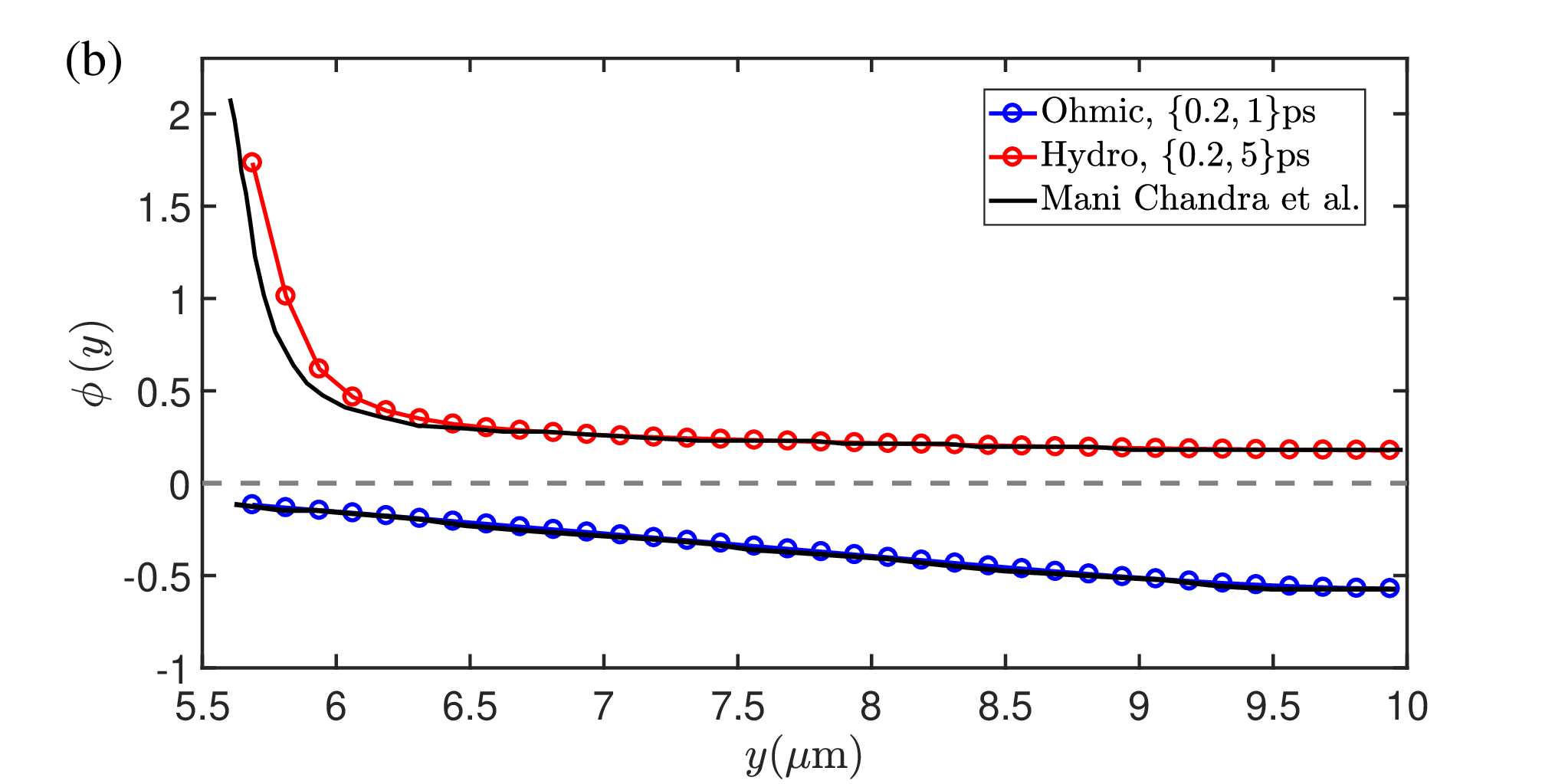}}~~
  \caption{\label{fig:8}(a) Change in normalized potential at $y = 90$ $\mu$m from $t = 200$ ps to $t = 300$ ps; (b) Phase difference of the potential waveform at $y\in \left[ 5.7,10 \right]$ $\mu$m with respect to $y = 50$ $\mu$m (i.e., the source current).}
\end{figure}

%\revision{Q: WHAT EXCATLY ARE THE DC AND AC DRIVING? TEMPERATURE, CHEMICAL POTENTIAL GRADIENT? OR ELSE?}
In this subsection, we calculate DC and AC electronic transport in a 2D sheet, using parameters of graphene. Assuming heavy $n$ type doping, we consider only the contribution from the upper band with the linear dispersion relation of $\varepsilon \left( k \right) ={\hbar }v_F {k}$, where $v_F \approx 10^6$ m/s is the Fermi velocity. Since hydrodynamic transport takes place at relative low temperatures, we set $T_0=10$ K, $\mu_0=10.8$ meV. We assume that DC and AC drive locate at the center of left and right boundaries with width of 1 $\mu$m. We adopt the same boundary conditions  as Ref.~\cite{PhysRevB.99.165409}. Specifically, for DC calculations, we set an isothermal boundary for the drift velocity $u_{x0}^{L}=u_{x0}^{R}=10^{-4}v_F$ distributed at the injections, and for AC calculations, we set an isothermal boundary for $u_{x0}^{L}=u_{x0}^{R}=10^{-4}v_F\sin \left( 2\pi ft \right) $, where $f = 10$ GHz is the frequency of AC driving. For the other boundaries, we use specular reflection. For such problems, one is often interested in the voltage or potential of the system. 
%which can be obtained by solving Poisson equation. However, when the external field is weak, Eq.~\ref{equ:1} is equivalent to the linear order of BTE containing the external force, i.e., $\nabla \bar{\mu}=\nabla \left( \mu -e \varphi \right) $, where $\varphi$ is the electric potential, and $\bar{\mu}$ is the electrochemical potential of the system, which can be obtained from our simulations (i.e., $\mu$ in the previous section).  Since the diffusion current contributed by $\nabla \mu$ is negligible due to the weak external field, we have
%\begin{equation}
%  \nabla \bar{\mu}=\nabla \mu -e\nabla \varphi \approx -e\nabla \varphi .
%  \label{equ:mu_V}
%\end{equation}
In our framework, we can obtain the potential by $\varphi \left( \boldsymbol{r,}t \right) =-{\mu}\left( \boldsymbol{r,}t \right) /e+\mu _0/e$.  We calculated the hydrodynamic transport dominated by the N-processes for the three cases with $\left\{ \tau ^N,\tau ^U \right\} =\left\{ 0.2, 5 \right\}$ ps, $\left\{ 0.2, 10 \right\}$ ps and $\left\{ 0.2, 50 \right\}$ ps, respectively. The results are shown in Figs.~\ref{fig:7a}-\ref{fig:7c}. Although all three sets of relaxation times are dominated by N-processes, only results from the third parameter set shows vortices. Positions of the vortices are consistent with the results of Mani \emph{et al}. \cite{PhysRevB.99.165409} and nonlocal negative resistance is also observed (Fig.~\ref{fig:7d}). 
%In other cases, \revision{they are only locally negative}, and the deviations close to the isothermal boundary are attributed to the potential jumps in this vicinity. Also, \revision{the local negative resistance} of the $\left\{ \tau ^N,\tau ^U \right\} =\left\{ 0.2, 50 \right\}$ ps case disappears when the number of meshes is further reduced, similar to the results of Mani et al.

For AC transport, we have considered two sets of parameters $\left\{ 0.2, 1 \right\}$ ps and $\left\{ 0.2, 5 \right\}$ ps. DUGKS again accurately captures the voltage transients (Fig.~\ref{fig:8a}). Fig.~\ref{fig:8b} shows that, for the Ohmic transport, the phase $\phi \left( y \right) =\phi \left[ I_d,\varphi \left( y \right) \right] $ follows the source current at any position, but for the hydrodynamic transport it consistently delayed from the source current.

\subsection{Thermoelectric transport}
\label{subsec:thermoelectric}
In this subsection, we consider thermoelectric transport in model metals and semiconductors with $L=200$ nm. The left and right chemical potentials are set to $\mu _L=\varepsilon _F+\Delta \mu/2$ and $\mu _R=\varepsilon _F-\Delta \mu/2$, where $\Delta \mu$ is 10 meV, 20 meV, and 40 meV, respectively. 
%For metals, the free electron approximation [EVEN IN METALS, IT IS THE EFFECTIVE MASS APPROXIMATION, SINCE ELECTRONS FEEL THE PERIODIC POTENTIAL.] is still considered, and
For semiconductors, we use the effective mass approximation $\varepsilon ={\hbar }^2{k}^2/2m^*$, with $m^*$ the electron effective mass. We assume that the semiconductor is  isotropic, $n$ doped with a conduction band effective mass $m^*=0.068m$. The Fermi energy level is located  $\Delta=0.05$ eV  below the conduction band bottom.
For both metal and semiconductor, the energy, real and angle space grids are set to $N_{\varepsilon}=80$, $N=80$, $N_{\theta}=100$, respectively. We considered both U ($\tau ^U=0.01$ ps, $\tau ^N=1$ s) and N ($\tau ^U=1$ s, $\tau ^N=0.01$ ps) dominated cases.  
%As shown in Fig.~\ref{fig:6a}, the green part is included in the calculation while the gray part does not contribute since it is in the band gap with zero density of states. 

%\revision{For both $\mu=10$ meV and 20 meV, the energy ranges are set to $\left[ \mu _R-12k_BT,\mu _L+12k_BT \right] $ and $\left[ \Delta ,\Delta +12k_BT \right] $ for metals and semiconductors, respectively; for $\mu=40$ meV, they are $\left[ \mu _R-15k_BT,\mu _L+15k_BT \right] $ and $\left[ \Delta ,\Delta +15k_BT \right] $.} 

%[PUT these parameters into the corresponding figure captions?]}

\begin{figure}[t]
  \centering
  \subfloat{\label{fig:5a}\includegraphics[width=0.34\textwidth]{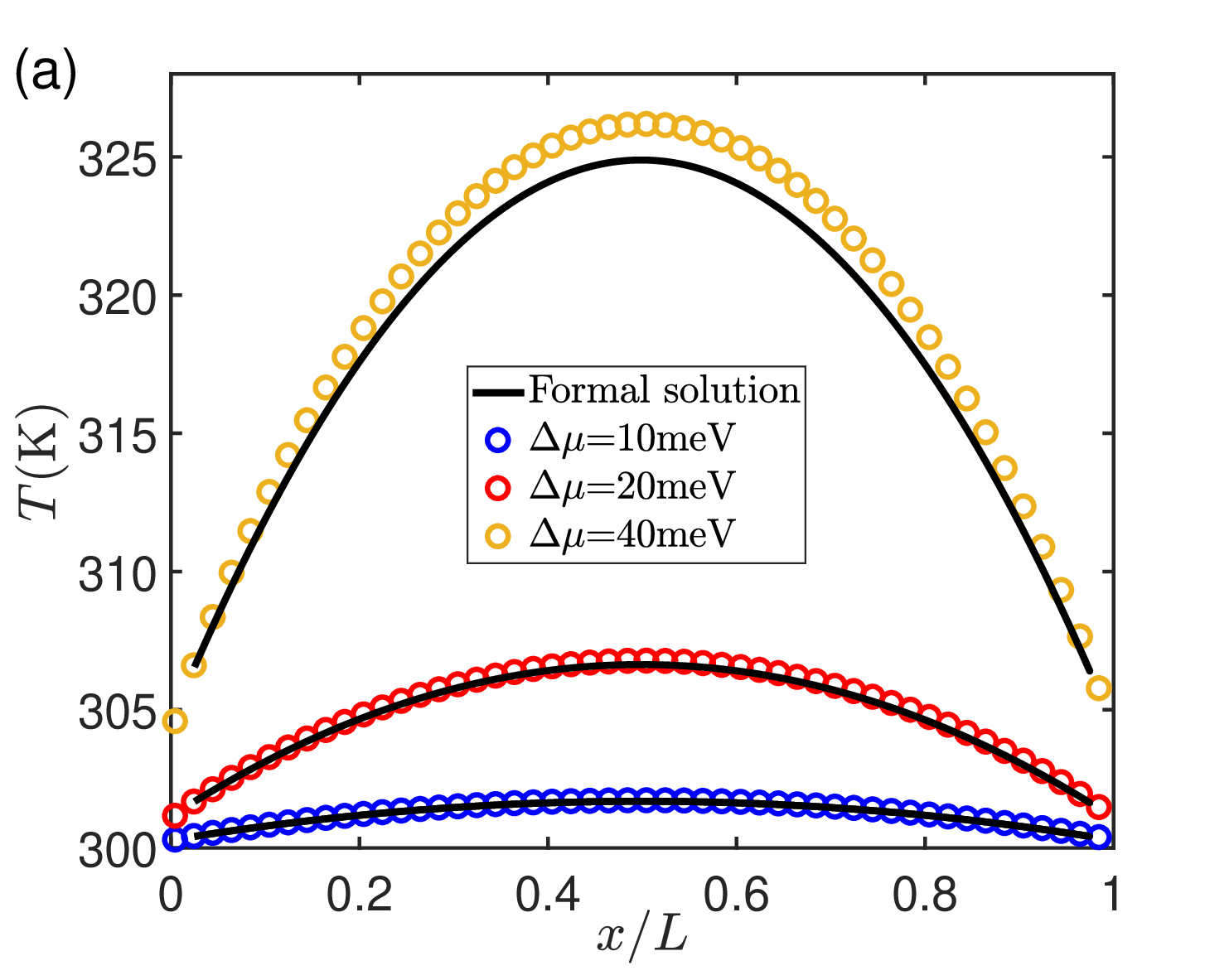}}~~
  \subfloat{\label{fig:5b}\includegraphics[width=0.34\textwidth]{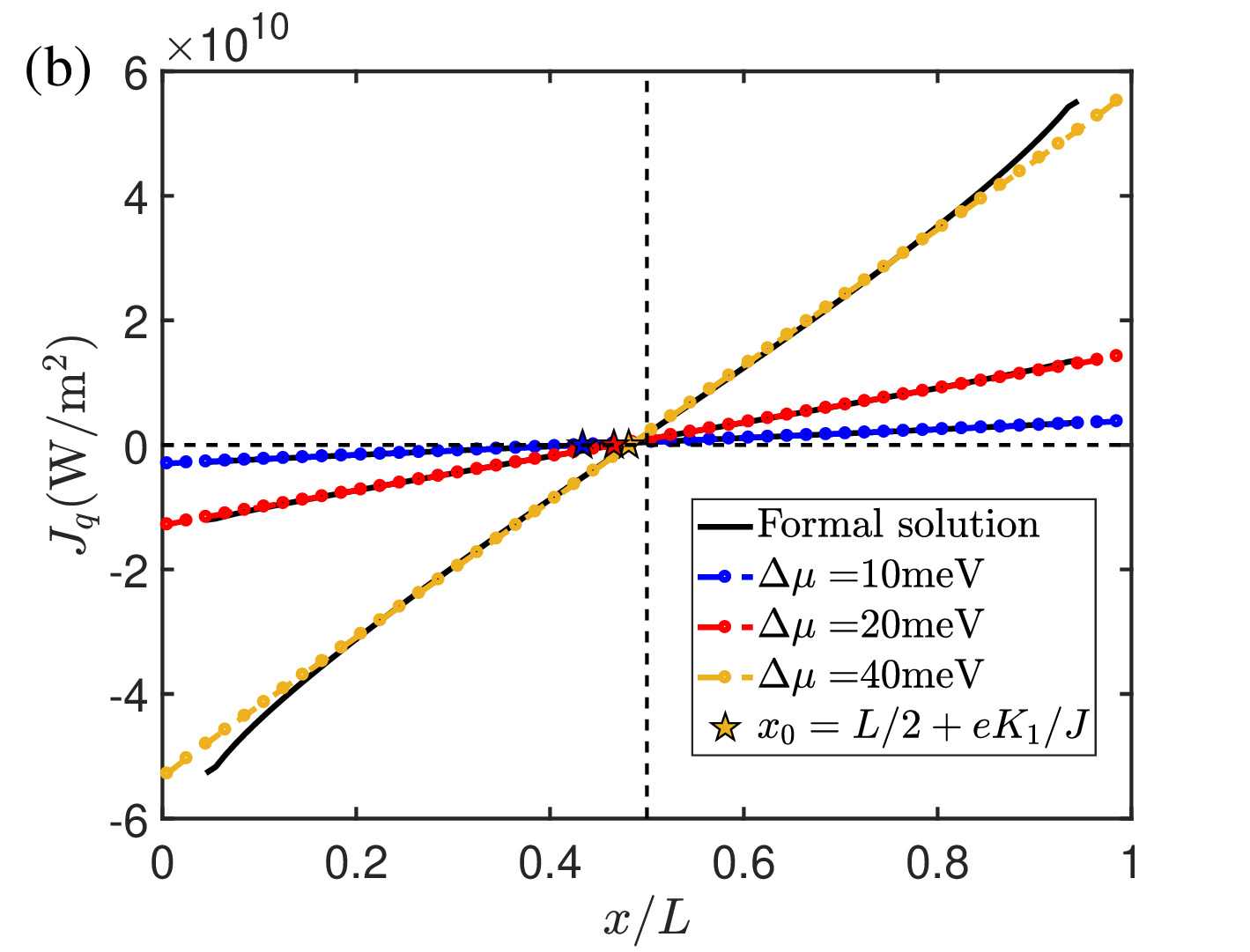}}~~\\
  \subfloat{\label{fig:5c}\includegraphics[width=0.34\textwidth]{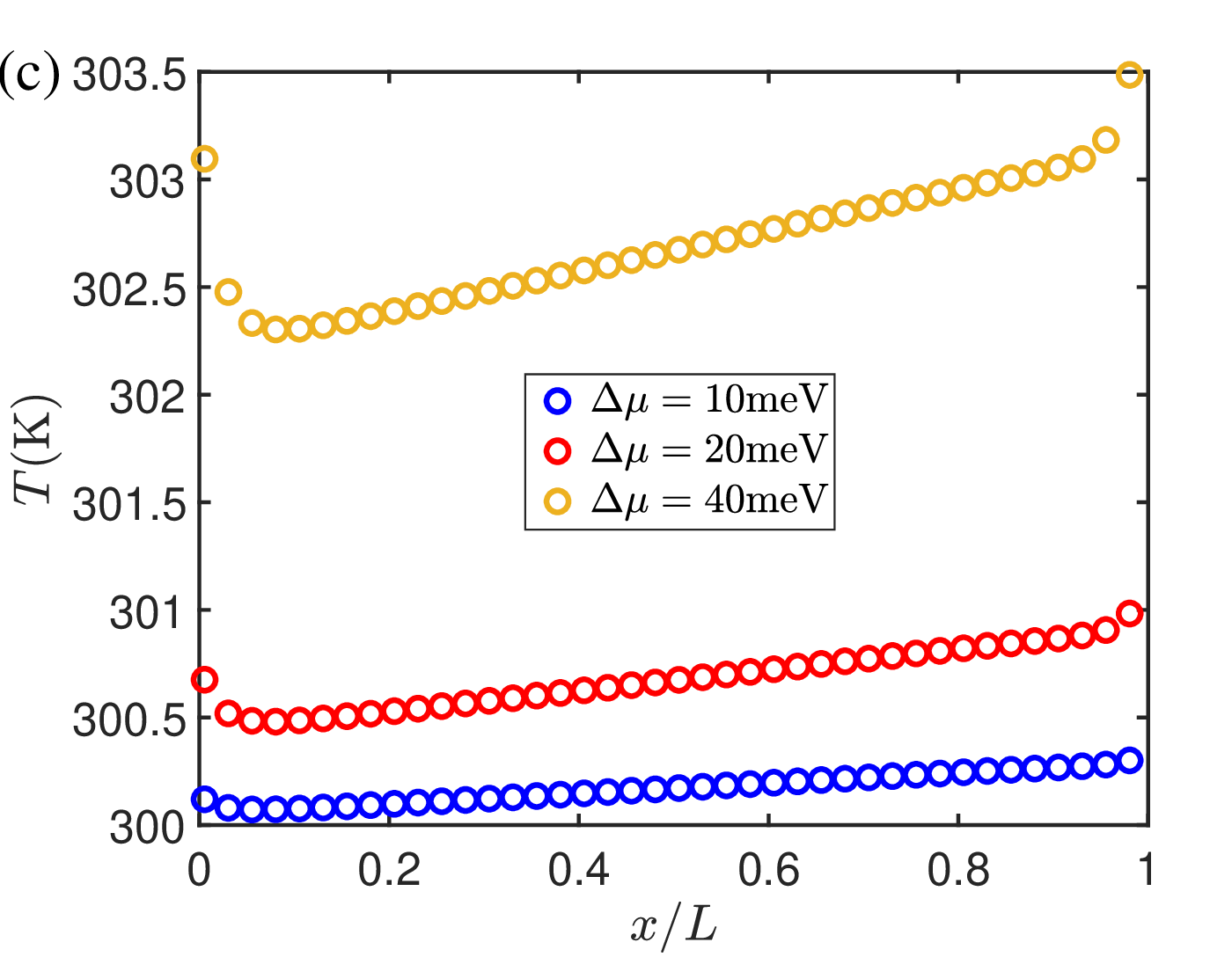}}~~
  \subfloat{\label{fig:5d}\includegraphics[width=0.34\textwidth]{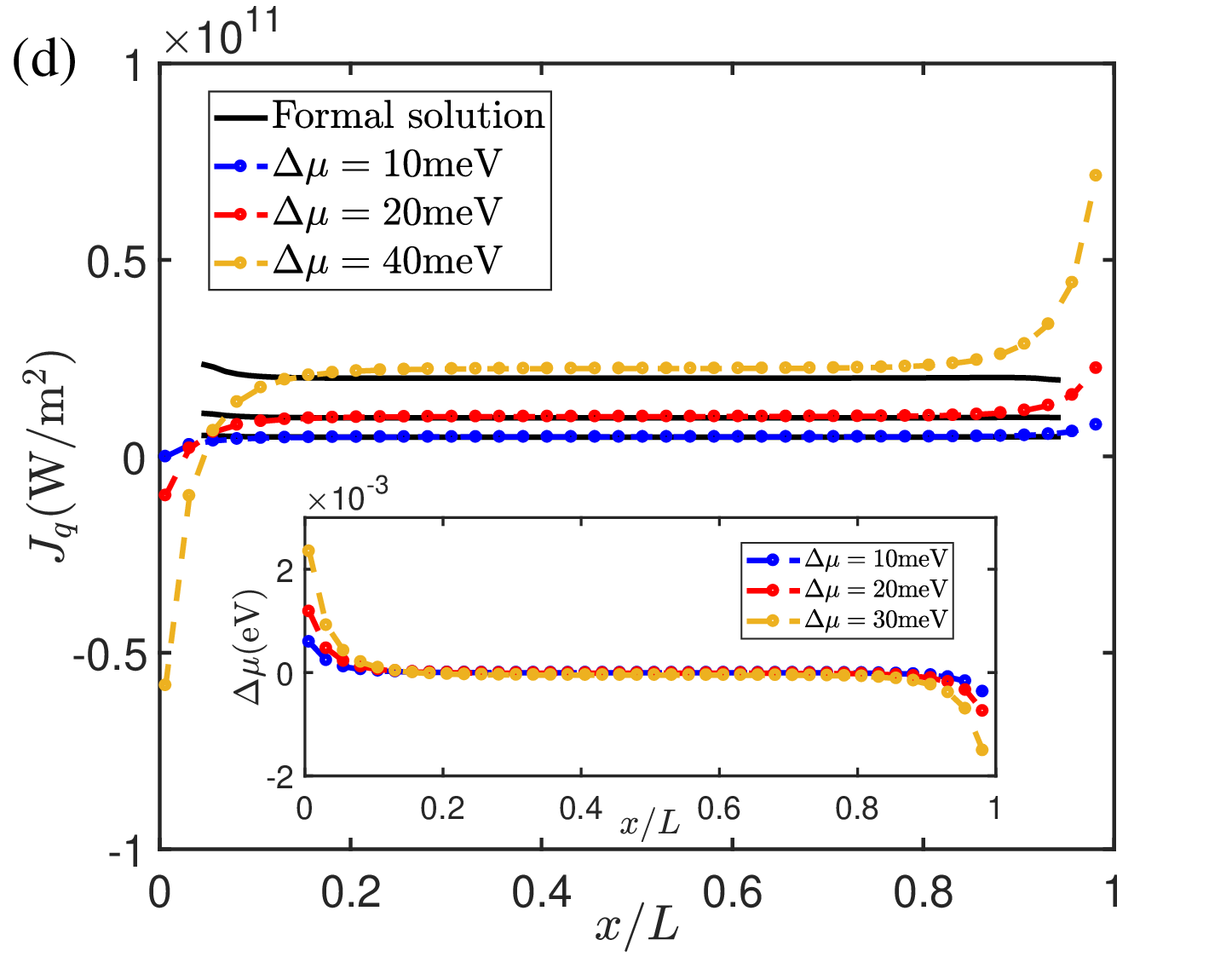}}~~
  \caption{\label{fig:5}Numerical results for metal. (a) Temperature distribution due to chemical potential difference for the U-process-dominated case. (b) The corresponding heat flux density distribution for the situation in (a). (c) Temperature distribution due to chemical potential difference in the N-process-dominated case. (d) The corresponding heat flux density distribution for the situation in (c). The inset shows the corresponding chemical potential distribution.}
\end{figure}

Although there is no analytical solution, we can interpret the results qualitatively using a first-order approximation of the eBTE. With the U-processes dominating, it is described as:
\begin{equation}
  \left( \begin{array}{c}
    \boldsymbol{J}\\
    \boldsymbol{J}_q\\
  \end{array} \right) =\left( \begin{matrix}
    eK_0\ \		-eK_1/T\\
    -K_1\ \		K_2/T\\
  \end{matrix} \right) \left( \begin{array}{c}
    \nabla \mu\\
    -\nabla T\\
  \end{array} \right) ,
  \label{equ:Onsager}
\end{equation}
and 
\begin{equation}
  K_n=\frac{\tau}{3}\iint{v}vD\left( \varepsilon \right) \left( \varepsilon -\mu \right) ^n\left( -\frac{\partial f_0}{\partial \varepsilon} \right) d\varepsilon d\Omega .
  \label{equ:Kn}
\end{equation}
In fact, Eq.~\ref{equ:Onsager} neglects the impact of boundary conditions, making it valid only in regions far from the boundaries \cite{PhysRevB.95.235137}. Thus, we describe the region of $x\in [$5, 195$]$ nm by Eq.~\ref{equ:Onsager}. 

The results for the metal in the U-process-dominated case are shown in Figs.~\ref{fig:5a}-\ref{fig:5b}. The temperature distribution takes the form of a quadratic function, with the peak occurring at $L/2$. This is expected when Joule heating is dominated \cite{PhysRevB.102.165134}.
In this case, using Eq.~\ref{equ:Onsager}, we can obtain a formal solution for the $T(x)$:
%The Seebeck coefficient is defined as $S=K_1/\left( eTK_0 \right)$, and the local heat flux is $\boldsymbol{j}_q=-K_1\nabla \mu -K_2\nabla \ln T$, which has an additional thermoelectric correction term. From Eq.~\ref{equ:Onsager}, we can obtain a formal solution for the temperature:
\begin{equation}
T\left( x \right) =-\frac{\bar{T}}{K_2}\frac{d\mu}{dx}\left( \frac{J}{e}x^2-\frac{J}{e}x_0x+K_1 x \right) +T_0+\Delta T_C,
  \label{equ:T}
\end{equation}
where $\Delta T_C$ is the temperature jump due to the thermal resistance of the interface, and $x_0$ is the zero point of heat flux. We neglect the spatial dependence of $K_n$ in obtaining Eq.~\ref{equ:T},
and assume that the chemical potential is linearly distributed. For metal, this applies away from the boundary. The results are shown in Fig.~\ref{fig:5a}, which agree well with the numerical results at low chemical potential differences. When $\Delta \mu = 40$meV, the large magnitude of the temperature change invalidates the above simple approximation. 
Now we consider the heat flux density in Fig.~\ref{fig:5b}.
We note the small deviation of $x_0$ at $L/2+eK_1/J$ from $L/2$, due to the thermoelectric correction term $eK_1/J$. As the chemical potential difference increases, the zero point shifts towards the center due to the increase in the current density. For $\Delta \mu = 40$ meV, the effect of the boundary conditions on the results is enhanced, leading to a deviation of the heat flux density close to the boundary.

%Meanwhile, the effect of the boundary condition results is enhanced, causing Eq.~\ref{equ:Onsager} to deviate from the BTE.
%we only show results within the region of $x\in [$5, 195$]$ nm in Fig.~\ref{fig:5a}, showing good agreement with numerical results.

\begin{figure}
  \centering
  \subfloat{\label{fig:6a}\includegraphics[width=0.34\textwidth]{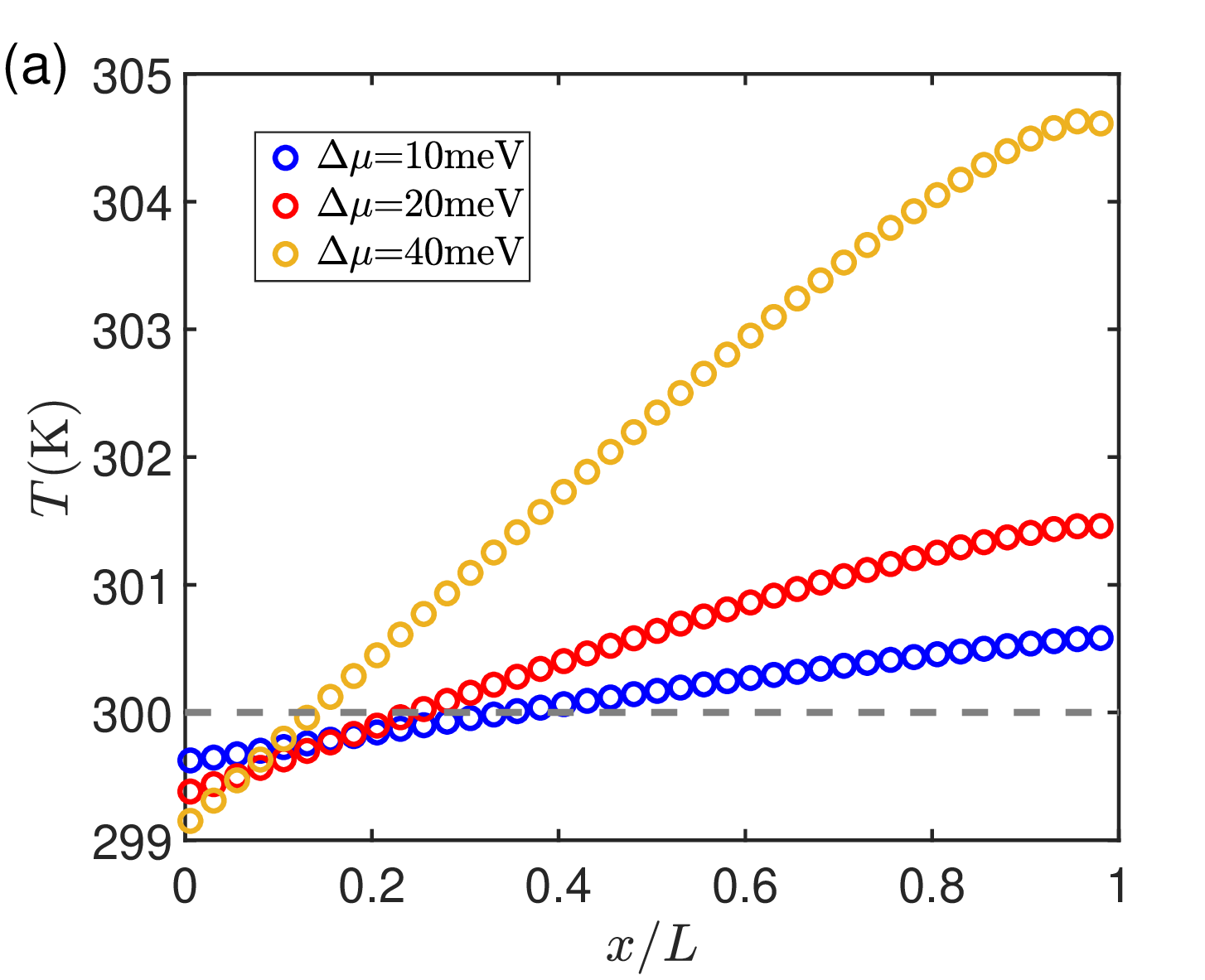}}~~
  \subfloat{\label{fig:6b}\includegraphics[width=0.34\textwidth]{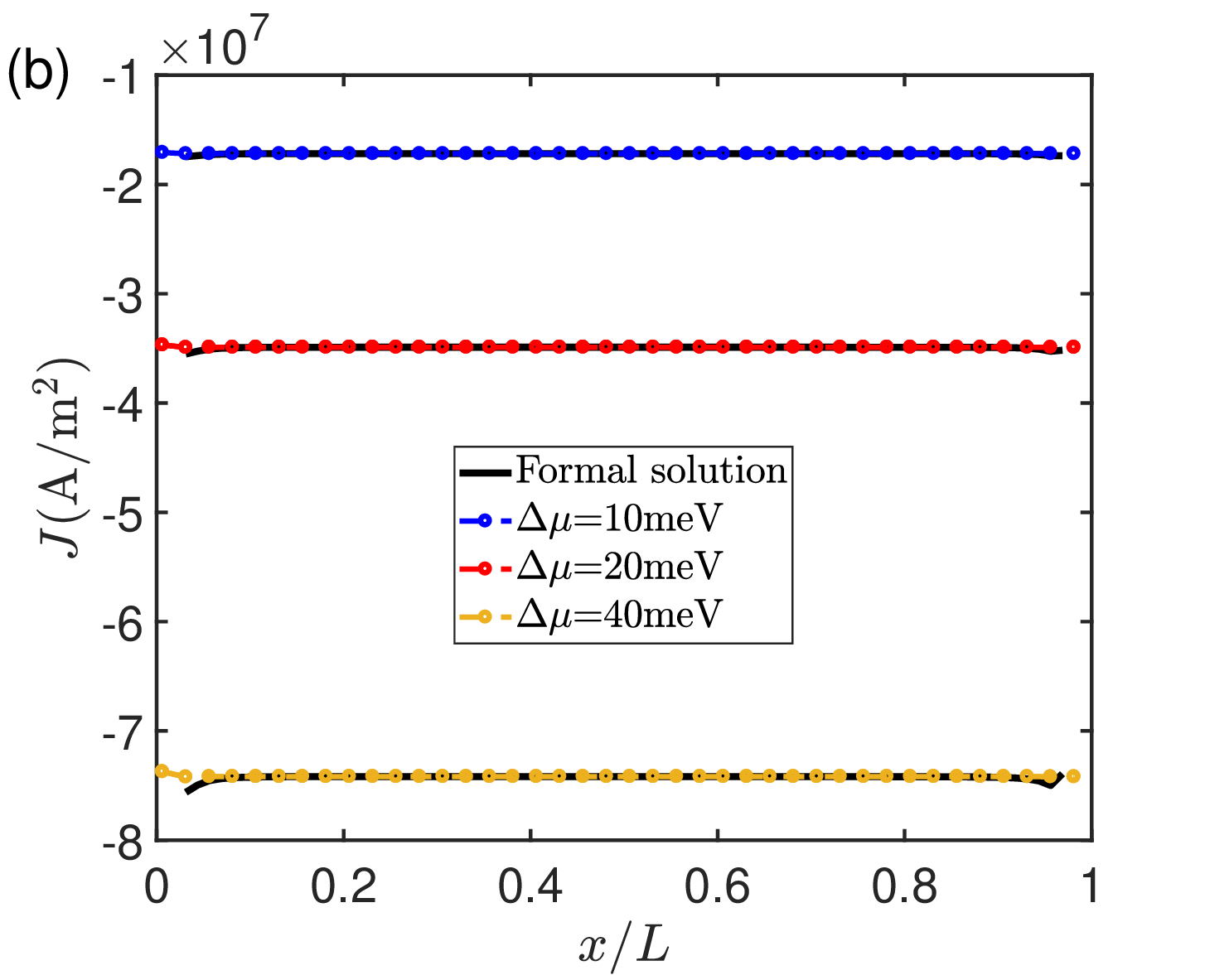}}~~\\
  \subfloat{\label{fig:6c}\includegraphics[width=0.34\textwidth]{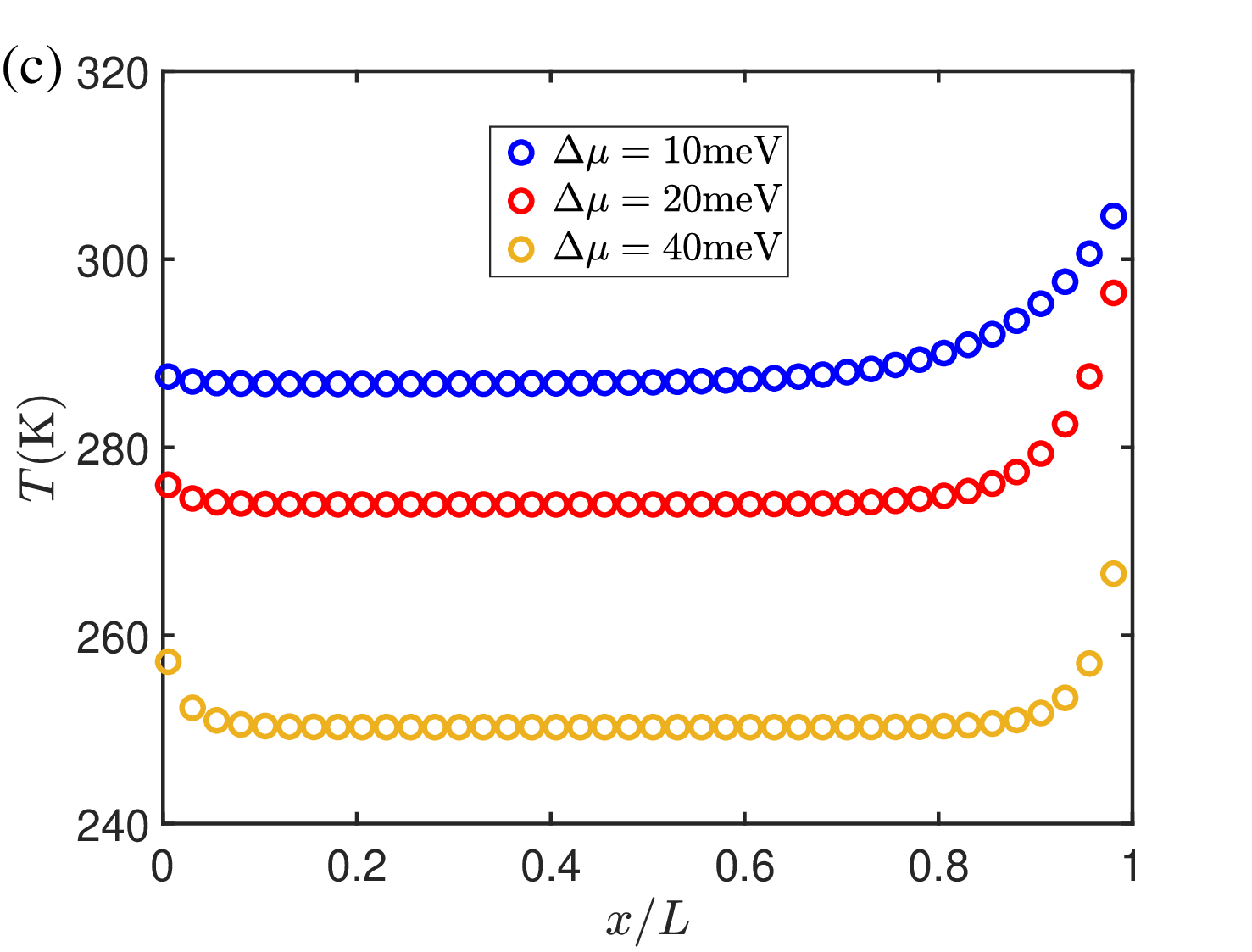}}~~
  \subfloat{\label{fig:6d}\includegraphics[width=0.34\textwidth]{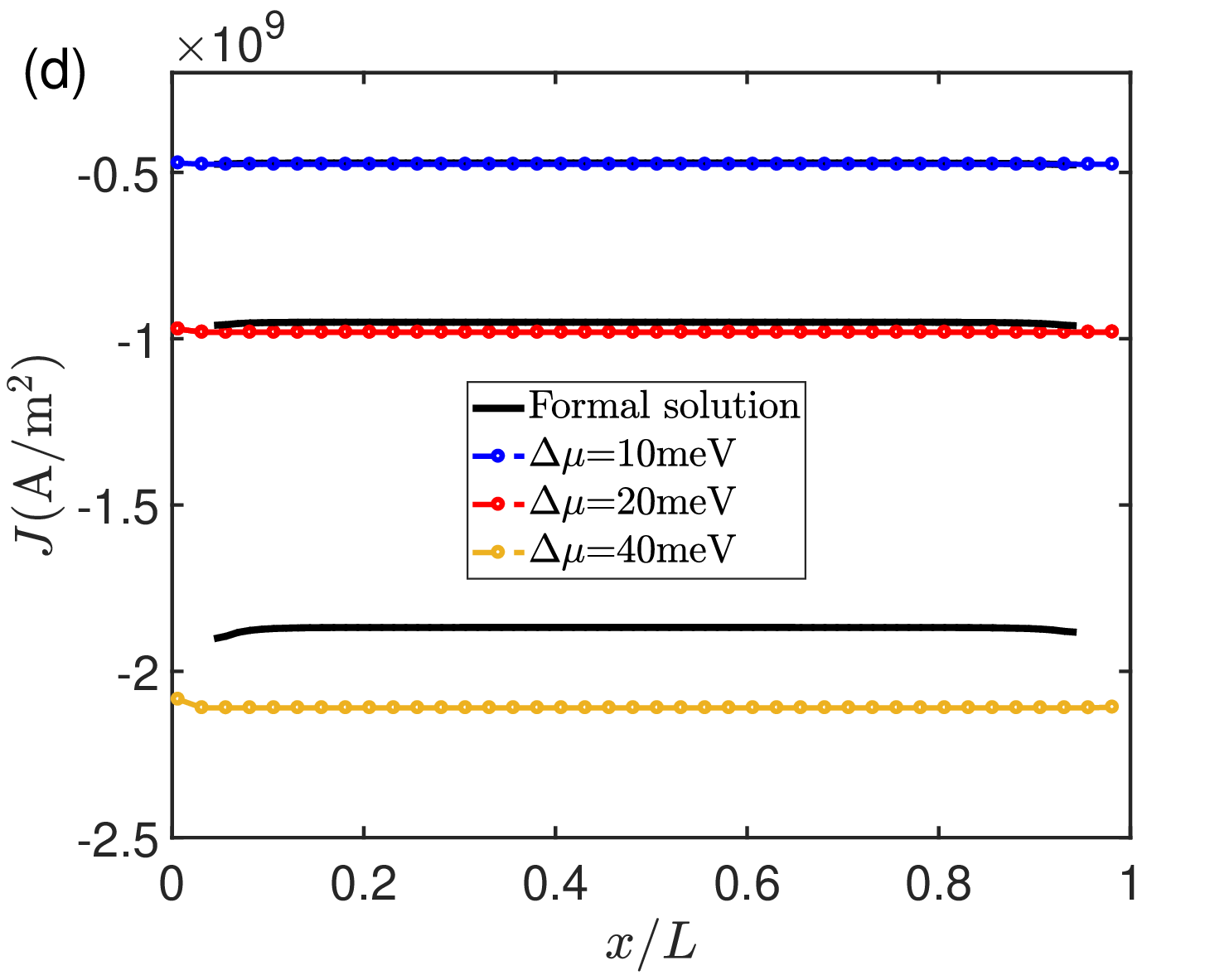}}~~
  \caption{\label{fig:6}Numerical results for semiconductor. (a) Temperature distribution due to chemical potential difference when the U-processes dominate. (b) The corresponding current density distribution for the situation in (a). (c) Temperature distribution in semiconductors due to chemical potential difference when the N-processes dominate. (d) The corresponding current density distribution for the situation in (c).}
\end{figure}

The results for the metal when N-process dominates are shown in Figs.~\ref{fig:5c}-\ref{fig:5d}. The presence of drift velocity $\boldsymbol{u}$ further complicates the results. Since the momentum conservation of the N-process does not generate thermal resistance, the temperature distribution should be homogeneous. However, the presence of boundary scattering makes the temperature non-uniform (See Fig.~\ref{fig:5c}). The electrons carry heat in from left boundary and do not dissipate heat in the middle. Eventually the electrons are scattered and release heat at the right boundary, resulting in higher temperatures on the right side than on the left. This non-uniformity becomes evident with the increase of the chemical potential difference. In addition, the temperature increase produced by the N-process is much lower than that of the U-process for the same chemical potential difference. The results of the heat flux are shown in Fig.~\ref{fig:5d}. They can be understood from the first-order approximation:

\begin{equation}
\left( \begin{array}{c}
	J\\
	J_q\\
\end{array} \right) =\left. \frac{1}{3}\iint{D}\left( \varepsilon \right) \left( \begin{array}{c}
	-e\\
	\varepsilon -\mu\\
\end{array} \right) \boldsymbol{v}\cdot \frac{\partial f_{0}^{N}}{\partial \boldsymbol{u}} \right|_{\boldsymbol{u}=0}d\varepsilon d\Omega \boldsymbol{u}+\left( \begin{matrix}
	eK_0&		-eK_1/T\\
	-K_1&		K_2/T\\
\end{matrix} \right) \left( \begin{array}{c}
	\nabla \mu\\
	-\nabla T\\
\end{array} \right) ,
  \label{equ:Onsager_N}
\end{equation}
where the first term on the right side is the effect of drift velocity. Since the temperature and chemical potential gradient are small, the contributions to the current and heat flux in the N-process come mainly from $\boldsymbol{u}$. The heat flux generated by the drift velocity term is nearly two orders of magnitude higher than the contributions from the temperature and the chemical potential. As in the previous analysis, momentum conservation makes the heat flux gradient zero in the region away from the boundary. The heat flux can also be written as $\boldsymbol{J}_q=\boldsymbol{J}_{\varepsilon}-\mu \boldsymbol{J}$, which satisfies $\nabla \boldsymbol{J}_{\varepsilon}=0$ and $\nabla \boldsymbol{J}=0$ at the steady state. This implies that the deviation of the heat flux from the first-order approximation arises mainly from the effect of boundary scattering on the chemical potential (See the inset of Fig.~\ref{fig:5d}).

Results for semiconductor model are depicted in Fig.~\ref{fig:6}. 
The temperature distribution when U-processes dominate is shown in Fig.~\ref{fig:6a}. Unlike metals, cooling takes place at the left boundary and becomes more apparent as the chemical potential difference increases. This is characteristic of the Peltier effect. The Peltier coefficient is defined as $\varPi =K_1/\left( eK_0 \right)$. From the numerical data, we get an average value of $\varPi=0.1$ V. Fig.~\ref{fig:6b} shows the corresponding current densities, which are consistent with Eq.~\ref{equ:Onsager}. 
For the case where N-processes dominate (Fig.~\ref{fig:6c}, \ref{fig:6d}), the conservation of momentum during the collision leads to the uniform decrease of temperature in the region away from the boundary. This cooling phenomenon is more pronounced compared to the U-process (Fig.~\ref{fig:6c}). The corresponding average Peltier coefficient is $\varPi= 0.09$ V, similar to the $U$-dominated case. However, the temperature decreases much more since the current density $\left| J \right|$ is two orders of magnitude larger than the case where the U-processes dominate (See Fig.~\ref{fig:6b} and \ref{fig:6d}).  The first-order approximation no longer holds as $\Delta \mu$ increases to 40 meV.

\section{Conclusion} \label{conclusion}
In summary, we developed a discrete unified gas kinetic scheme for the solution of electron Boltzmann transport equation under Callaway approximation. The coupled treatment of electron drift and scattering makes the cell size and time step independent of the mean free path and relaxation time, which has advantage in the study of problems with small Knudsen number. Numerical results demonstrate that the scheme accurately captures electron transport behaviors across ballistic, hydrodynamic, and diffusive regimes, while also exhibiting asymptotic preservation properties. Due to the consideration of the electronic energy band structure and the use of the Newtonian method to solve the energy and particle number conservation equations at the cell interfaces and centers, we are able to simulate different materials across a wide range of parameter regimes. 
%In metals, heat is generated as Joule heat under the driving of a chemical potential difference during the momentum-relaxation processes. On the other hand, in semiconductors, the thermoelectric effect is more pronounced, resulting in a decrease in temperature. This feature is particularly prominent in N-processes due to higher current densities that carry away more heat compared to U-processes.
%
Meanwhile, more complex device shapes, more realistic energy band structure can also be incorporated into our framework in the future. Study of coupled transport including more than one type of quasiparticles, i.e., including photon-electron-phonon coupled system, is also possible under this generic scheme. 
However, the dynamics of electrons in $k$-space needs to be included to study transport under strong electric and magnetic fields. This issue will be discussed in the future.

%This work was supported by the National Natural Science Foundation of China (22273029, 51836003) and the Interdiciplinary Research Program of HUST (2023JCYJ002).





%\begin{thebibliography}{00}



%\end{thebibliography}
\biboptions{numbers,sort&compress,square,comma}
\bibliographystyle{elsarticle-num} 
\bibliography{paper} 
\end{document}